\documentclass[journal]{IEEEtran}
\usepackage{bbm}
\usepackage{amsmath,amssymb}
\usepackage{amsfonts}
\usepackage{stfloats}

\usepackage{amsbsy}
\usepackage{graphics}
\usepackage{graphicx}
\usepackage{float}

\usepackage{subfloat}
\usepackage{epstopdf}
\usepackage{chngcntr}
\usepackage{wrapfig}
\usepackage{caption}
\usepackage{xcolor}

\newtheorem{thm}{Theorem}

\newtheorem{lemma}{Lemma}

\newtheorem{remark}{Remark}

\usepackage{mathrsfs}
\usepackage{dsfont}
\usepackage{yfonts}
\usepackage{bbm}
\usepackage{scalerel}

\usepackage{cuted}
\usepackage[linesnumbered]{algorithm2e}
\RestyleAlgo{ruled}

\makeatletter
\g@addto@macro{\@algocf@init}{\SetKwInOut{Parameter}{Parameters}}
\makeatother

\usepackage{color, colortbl}
\definecolor{dark-gray}{gray}{0.68}
\definecolor{light-gray}{gray}{0.9}

\usepackage{multirow}

\usepackage[colorlinks,bookmarksopen,bookmarksnumbered,citecolor=blue,urlcolor=blue]{hyperref}

\usepackage{amscd}
\usepackage{mathrsfs}

\usepackage{etoolbox}

\makeatletter
\newif\ifhb@open
\hb@openfalse

\let\hb@orig@bibitem\@bibitem
\def\@bibitem#1{%
  \ifhb@open\endgroup\fi
  \hb@opentrue
  \begingroup
  \ifcsdef{hb@#1}{\color{\csuse{hb@#1}}}{}%
  \hb@orig@bibitem{#1}\ignorespaces
}

\let\hb@orig@lbibitem\@lbibitem
\def\@lbibitem[#1]#2{%
  \ifhb@open\endgroup\fi
  \hb@opentrue
  \begingroup
  \ifcsdef{hb@#2}{\color{\csuse{hb@#2}}}{}%
  \hb@orig@lbibitem[#1]{#2}\ignorespaces
}

\pretocmd{\endthebibliography}{\ifhb@open\endgroup\hb@openfalse\fi}{}{}
\makeatother

\usepackage{cite}

\usepackage{amsmath,amssymb,mathtools,bm}
\usepackage{physics}
\usepackage{bbm}
\usepackage{graphicx}

\newcommand{\T}{{\scriptscriptstyle\mathrm{T}}}
\renewcommand{\H}{{\scriptscriptstyle\mathrm{H}}}

\usepackage{caption}
\usepackage{subcaption}

\begin{document}
\title{UAV-Mounted Hybrid RIS-Assisted ISAC-RSMA: Performance Analysis with Imperfect CSI and SIC 
}
\author{{Chandan~Kumar~Singh}, \IEEEmembership{Member,~IEEE},
{Arthur~Sousa~de~Sena}, \IEEEmembership{Member,~IEEE},
{Smriti~Uniyal}, \IEEEmembership{Graduate Student Member,~IEEE},
{Deepak~Kumar}, \IEEEmembership{Member,~IEEE},  {Prabhat~K.~Upadhyay}, \IEEEmembership{Senior Member,~IEEE}, and {Matti Latva-aho}, \IEEEmembership{Fellow,~IEEE}
		  \thanks{Copyright (c) 20xx IEEE. Personal use of this material is permitted. However, permission to use this material for any other purposes must be obtained from the IEEE by sending a request to pubs-permissions@ieee.org.}
	  \thanks{This work was supported in part by Business Finland, Nokia, Keysight, and Finwe via 6G-WISECOM project.}
	 \thanks{C. K. Singh was with the CWC, University of Oulu, Finland. He is now with the Department  of Electronics and Communication Engineering, Dr. B.R. Ambedkar National Institute of Technology Jalandhar, India (email: singhck@nitj.ac.in). A. S. d. Sena is with the Ericsson, Sweden (e-mail: arthurssena@ieee.org). S. Uniyal and M. Latva-Aho are with the CWC, University of Oulu, Finland (e-mail: \{ smriti.uniyal, matti.latva-aho\}@oulu.fi). D. Kumar is with the Department of Electronics and Communication Engineering, Motilal Nehru National Institute of Technology Allahabad, India (e-mail: dkumar@mnnit.ac.in). P. K. Upadhyay is with the Department of Electrical Engineering, IIT Indore, India (e-mail: pkupadhyay@iiti.ac.in).}
\vspace*{-0.50cm}}
\maketitle
\begin{abstract}\label{abst}
In integrated sensing and communication (ISAC) systems, strict sensing requirements can substantially constrain the transmit power available for communication. A hybrid reconfigurable intelligent surface (HRIS) with passive reflection and active amplification capabilities mounted on an unmanned aerial vehicle (UAV) can enhance coverage and strengthen communication links in such power-limited regimes. Meanwhile, rate-splitting multiple access (RSMA) provides an effective transmission strategy for managing inter-user interference, particularly in dense multiuser wireless scenarios. Motivated by this, we analyze the performance of a UAV-mounted HRIS-assisted monostatic ISAC system employing RSMA scheme. We consider the imperfections in the channel state information and successive interference cancellation in the RSMA, which are inevitable in practical system implementations. We then derive highly accurate expressions for outage probability, ergodic rate, and throughput to evaluate communication performance over Nakagami-$m$ fading channels. These expressions reveal key insights into the diversity order and coding gain of the system in the high signal-to-noise ratio regime. To characterize the sensing performance, we derive expressions for the probability of detection, the probability of false alarm, and the Cramér-Rao lower bound. Numerical results validate the theoretical findings and highlight the performance gains of the proposed system over conventional multiple-access schemes and RIS-based systems.
\end{abstract}
\begin{IEEEkeywords}
Hybrid reconfigurable intelligent surface (HRIS), integrated sensing and communication (ISAC), mobile networks, rate-splitting
multiple access (RSMA), and unmanned aerial vehicle (UAV) communication.
\end{IEEEkeywords}

\section{Introduction}\label{Intro}
\IEEEPARstart{N}{ext}-generation wireless technologies, such as evolving 6G systems, are expected to meet increasingly stringent demands. These demands include high throughput, low latency, improved spectral efficiency, and supporting massive connectivity. In addition, 6G is envisioned to integrate sensing capabilities with communication to enable reliable situational awareness under shared time-frequency resources \cite{gonzalez2025six}. This need has driven the rise of integrated sensing and communication (ISAC) systems, which jointly deliver data transmission and radar-like sensing using a unified waveform and infrastructure \cite{yang2024integrated}. However, achieving reliable communication and sensing in practical deployments is challenging due to fading, blockages, and multiuser interference, and because the sensing requirement can significantly constrain the transmit power available for communication \cite{zhu2024enabling}.

To address these challenges in 6G mobile networks, emerging technologies such as reconfigurable intelligent surfaces (RISs), unmanned aerial vehicles (UAVs), and advanced multiple-access schemes have gained significant attention \cite{pan2021reconfigurable,liu2023integrated,nguyen2023fairness,clerckx2023primer}. RIS technology electronically controls wireless propagation using low-cost reconfigurable elements to enhance signal quality and coverage \cite{pan2021reconfigurable}. Passive RIS (PRIS) can optimize propagation but suffers from double path loss, whereas active RIS (ARIS) amplifies the impinging signals at the expense of additional power consumption and noise \cite{liu2023integrated}. A hybrid RIS (HRIS) combines active and passive elements to balance signal enhancement and energy efficiency, where even a small number of active elements can significantly outperform a fully PRIS. When mounted on UAVs, an HRIS can be repositioned to reinforce weak links and bypass obstacles in harsh environments \cite{nguyen2023fairness}. To mitigate blockage and coverage degradation, controllable propagation and flexible deployment are required in ISAC-enabled mobile networks. In this context, UAV-mounted hybrid RIS offers on-demand LoS enhancement and link reconfiguration. Efficient multiple access is also essential for dense user deployments, orthogonal multiple access (OMA) is spectrally limited, while non-orthogonal multiple access (NOMA) is constrained by rigid power-domain allocation and interference propagation. Rate-splitting multiple access (RSMA) addresses these limitations by splitting messages into common and private components, enabling robust interference management and improved spectral efficiency under practical imperfections \cite{clerckx2023primer}.

\subsection{Related Works}
ISAC has gained increasing attention as a key enabling technology for 6G mobile networks, since it unifies spectrum usage and radio infrastructure to jointly provide high-rate communication and radar-like sensing services \cite{ye2024general}. Accurate propagation characterization is a critical prerequisite for performance evaluation and resource allocation in ISAC. Nguyen {\it{et al}.} \cite{nguyen2025performance} derived the closed-form expressions for the communication rate and sensing Cramér–Rao lower bound (CRLB) for massive multiple-input-multiple-output (MIMO) ISAC system. A stochastic model in \cite{gan2024coverage} showed the sensing benefits for communication. On the other hand, a blockage-aware ISAC model was studied in \cite{sun2024performance} and the performance tradeoffs between sensing and communication were analyzed in \cite{ouyang2023mimo}.

Recently, RIS technology has attracted attention for enhancing wireless propagation via programmable reflections, and several studies have analyzed its performance in various setups \cite{vu2022performance,zhang2022active,nguyen2024performance,li2023performance,yue2025astars}. Vu \textit{et al.} \cite{vu2022performance} analyzed full-duplex RIS under imperfect channel state information (iCSI) and quantized phase shifts, highlighting practical performance limitations and enhancements. In \cite{zhang2022active,nguyen2024performance}, ARIS was introduced to overcome the limitations of PRIS, showing notable capacity gains via signal amplification. A simultaneously transmitting and reflecting RIS (STAR-RIS) cognitive radio NOMA system was studied in \cite{li2023performance,yue2025astars} for reliability and efficiency. STAR-RIS-assisted NOMA systems were studied in \cite{li2024covert,zhang2024starris,guo2023star}, demonstrating improved outage and delay performance under hardware impairments, imperfect successive interference cancellation (iSIC), and Rician fading in non-terrestrial networks. A hybrid relay-reflecting RIS was shown in  \cite{nguyen2022spectral} to enhance spectral efficiency in cell-free massive MIMO systems. In \cite{vu2024hybrid} \cite{uniyal2025outage}, hybrid active-passive RIS NOMA systems were analyzed for outage and energy efficiency under fading and iSIC conditions. Ntougias \textit{et al.} \cite{ntougias2025hybrid} proposed HRIS and sub-connected active RIS designs to enhance capacity and energy efficiency under practical power constraints.  With respect to ISAC integration, Yu \textit{et al.} \cite{yu2024irs} proposed a RIS-aided NOMA-ISAC system with joint beamforming. ARIS has been shown to outperform PRIS in both communication and sensing  \cite{parihar2025performance}. RIS-UAV systems improved throughput under iCSI \cite{singh2022noma,singh2023performance}, while \cite{ghosh2024uav} demonstrated performance gains in disaster scenarios via diversity and ergodic capacity analysis. An RIS-UAV-NOMA system showed improved outage and spectral efficiency over the baseline schemes \cite{Solanki_TVT}. Deng \textit{et al.} \cite{deng2024joint} proposed a joint UAV beamforming-trajectory optimization for maximizing covert rate in ISAC networks. 

{
RSMA has emerged as a robust multiple-access and interference-management framework for future wireless networks~\cite{clerckx2023primer,mao2022rate}. Aditya \textit{et al.} investigated RSMA for 6G using link- and system-level simulations and experimental prototypes, showing its performance gains over spatial division multiple access (SDMA)~\cite{aditya2025rate}. In contrast, Vu \textit{et al.} showed that RSMA achieves a lower outage probability and a higher ergodic rate than NOMA~\cite{vu2023performance}.
Recent works have also investigated the integration of RSMA with RISs. Narottama \textit{et al.} investigated RIS-assisted RSMA and maximized energy efficiency using a quantum-machine-learning-based optimization framework~\cite{narottama2024qml}. Chang \textit{et al.} studied STAR-RIS-aided MISO-RSMA systems and focused on maximizing the rate~\cite{chang2024covert}. Furthermore, the use of RSMA with RIS has been explored for  covert performance~\cite{liang2025covert}, while the performance of STAR-RIS-assisted RSMA systems was analyzed in~\cite{Z_Junyao}.
More recently, the integration of RSMA with ISAC has attracted increasing attention. Chen \textit{et al.} studied RSMA for simultaneous multi-user communication and multi-target sensing and optimized the tradeoff between the max-min fairness rate and the sensing CRLB~\cite{chen2024multiuser}. Liu \textit{et al.} investigated STAR-RIS-enabled ISAC systems with RSMA by jointly optimizing rate splitting, transmit beamforming, and passive beamforming under communication and sensing requirements~\cite{liu2025starris}. Lyu \textit{et al.} experimentally evaluated RSMA-assisted ISAC using sum throughput and radar SNR as performance metrics, showing practical gains over SDMA~\cite{lyu2026experimental}. In addition, RSMA was shown to enhance interference mitigation in ISAC systems~\cite{ghosh2024performance}, while RIS-aided ISAC-RSMA frameworks with joint beamforming and phase-shift optimization were studied in~\cite{chen2025optimization,zhang2024robust}. An RIS-assisted ISAC-RSMA system was also studied in~\cite{11048955}, where energy efficiency was addressed via reinforcement learning in blockage-prone environments.}

\subsection{Motivations and Key Contributions} Despite the growing interest in RIS-assisted ISAC with RSMA, the potential of HRIS in such frameworks remains largely underexplored. The existing works on RIS-assisted RSMA based ISAC systems considered PRISs \cite{chen2025optimization}. However, the resulting monostatic four-hop echo path due to the PRIS, i.e., BS-RIS-target-RIS-BS, can lead to severe cascaded path loss, yielding a weak echo at the BS. Furthermore, many existing works did not account for practical impairments, including iCSI, residual SIC interference, and fading effects \cite{li2023performance, li2024covert, Solanki_TVT, Z_Junyao}, which are unavoidable in practical deployments. 
In \cite{ghosh2024performance}, closed-form expressions for the rate and outage of an RSMA-based ISAC system were presented. However, their insights were limited to the case of a single-antenna BS. \cite{deng2024joint} focused on a UAV beamforming optimization in ISAC networks. Additionally, existing works have mainly focused on demonstrating the effectiveness of RIS-assisted RSMA in enhancing the communication rates \cite{liang2025covert}. In contrast, the analytical characterization of sensing performance metrics such as the probability of detection within RIS-assisted RSMA-ISAC frameworks remains relatively limited and is still in a nascent stage. Such analysis is crucial to quantify sensing reliability in closed form and clarify how system parameters and resource allocation can affect the detection performance. 

Motivated by the aforementioned research gaps, this work investigates the performance of an HRIS-assisted monostatic ISAC system with an RSMA transmission scheme to enable robust wireless connectivity and accurate sensing in multiuser wireless scenarios. In blockage-prone wireless environments dominated by non-line-of-sight (NLoS) propagation, a UAV-mounted HRIS can enhance coverage and provide reliable connectivity for ground users. Moreover, RSMA can support dense multiuser deployments by flexibly managing the inter-user interference. Our contributions are summarized as follows:
\begin{itemize}
\item We propose a UAV-mounted HRIS-assisted monostatic ISAC system employing RSMA scheme at a multi-antenna source to serve multiple ground users while sensing a target in challenging NLoS scenarios. We derive analytical expressions for the signal-to-interference-plus-noise ratios (SINRs) by considering realistic CSI imperfections and residual interference due to SIC errors. This overcomes the limitations of prior works that assume perfect CSI knowledge and perfect successive interference cancellation (pSIC).
\item Leveraging the derived SINRs, we provide highly accurate analytical expressions for key communication performance metrics, namely outage probability (OP), ergodic rate (ER), and throughput, under both delay-limited (DL) and delay-tolerant (DT) transmission modes.  Furthermore, we conduct asymptotic analysis in the high signal-to-noise ratio (SNR) regime to provide useful insights into the diversity order and coding gain, highlighting their dependence on CSI accuracy, the number of transmit antennas, and the Nakagami-$m$ fading parameters. To the best of our knowledge, analytical results for the considered system over Nakagami fading channels have not been reported in the literature.
\item To characterize the sensing performance of the proposed ISAC system, we derive expressions for the probability of detection (PoD), the probability of false alarm (PoFA), and the CRLB. Our analysis demonstrates that increasing the number of transmit antennas and the sensing power allocation coefficient enhances sensing performance, as reflected by the reduced CRLB values.
\item Using extensive Monte Carlo simulations, we validate all the derived analytical results. The numerical findings highlight that (i) a small number of active elements in the HRIS  significantly enhances the performance compared to the PRIS-assisted system, (ii) increasing the number of HRIS elements and transmit antennas substantially improves the OP and ER. However, at high SNR, the ER saturates owing to interference limitations, and (iii) the OP is maximized when the UAV is at the farthest point of its trajectory from the users and minimized when it is closest. Moreover, increasing the trajectory radius results in a higher OP when the UAV is far from the users, whereas it results in a lower OP when the UAV is close to the users

\item {We then investigate the tradeoff between} the sensing and communication performance by analyzing the impact of the sensing power allocation coefficient on the CRLB and PoD. The results demonstrate that increasing the sensing power allocation coefficient leads to a reduced CRLB and a higher probability of detection. However, it also increases the OP and reduces the ER, highlighting the communication--sensing tradeoff in ISAC systems.
\end{itemize}
\subsection{Organization and Notations}
The rest of the paper is structured as follows. Section~\ref{system_model} details the communication and sensing models used in this work. Section~\ref{performance} analyzes the communication metrics, including OP, ER, and throughput. Section~\ref{performance_Sensing} provides the expressions for the sensing metrics, such as the PoD, PoFA and CRLB. The simulation results are presented in Section~\ref{num}, and the conclusions are presented in Section~\ref{conclu}.

Throughout this paper, scalars, vectors, and matrices are denoted by lowercase, bold lowercase, and bold uppercase letters, respectively. The operators $\mathbb{E}[\cdot]$, $\mathrm{Var}[\cdot]$, $|\cdot|$, and $\|\cdot\|$ represent expectation, variance, modulus, and Euclidean norm, respectively. We use $\bar{\mathrm{H}}^{\cdot,\cdot}_{\cdot,\cdot}\!\left[\cdot \mid \cdot \right]$ for the extended Fox-H function, $\gamma(\cdot,\cdot)$ for the lower-incomplete Gamma function, $\mathcal{CN}(\cdot,\cdot)$ for the complex Gaussian distribution, $\mathrm{Nak}(\cdot,\cdot)$ for the Nakagami-$m$ distribution, and $\mathrm{Gamma}(\cdot,\cdot)$ for the Gamma distribution.

\section{System Model}\label{system_model}
 \begin{figure}[t!]
	\centering
\includegraphics[height=7cm,width=8.5cm]{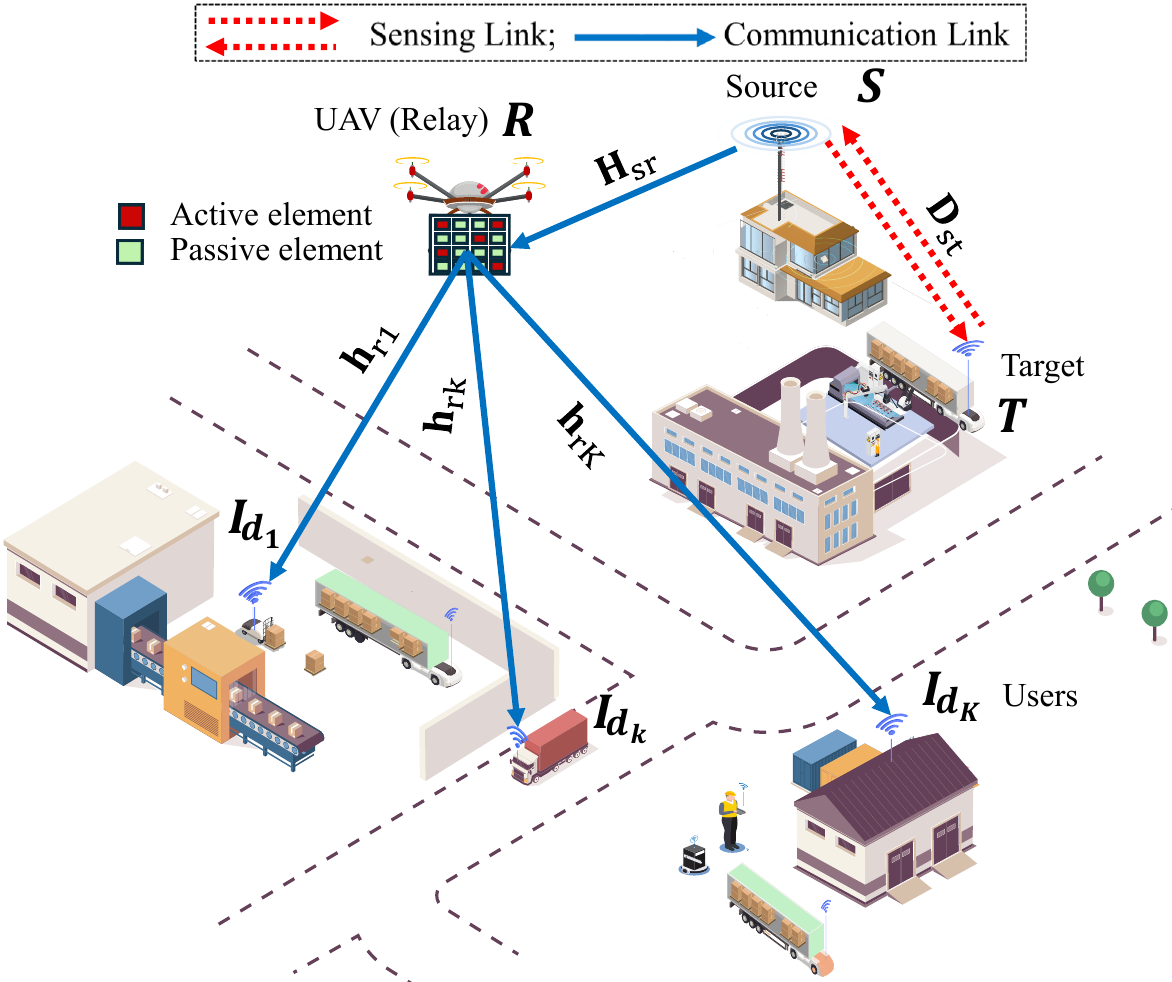}
	\caption{\small UAV-mounted HRIS-assisted monostatic ISAC system employing RSMA scheme.}
	\label{Schematic_diagram}
\end{figure}

Fig.~\ref{Schematic_diagram} illustrates a UAV-mounted HRIS-assisted monostatic ISAC system employing RSMA transmission scheme. Specifically, the system comprises a multi-antenna source ($S$), an HRIS mounted on a UAV (HRIS-UAV) acting as a relay ($R$), $K$ single-antenna users ($I_{d_k}, k=1, \dots,K$), and a single sensing target ($T$). The source is equipped with $N_{\mathrm{t}}$ transmit and $N_{\mathrm{r}}$ receive antennas, whereas the HRIS-UAV operates as an active/passive relay with a total of $N$ reconfigurable elements. {In the considered blockage-prone multiuser wireless scenario, the ISAC-enabled source
simultaneously transmits data signals to the ground users and probing signals to a target $T$ located
at a specific angle of interest. This angle is assumed to have been determined during the radar
scanning or searching phase \cite{nguyen2025performance}.} The HRIS-UAV enhances coverage and link reliability by providing flexible deployment, which is particularly beneficial for obstructed environments with NLoS users. On the other hand, RSMA enables multiuser transmission by flexibly managing inter-user interference under practical impairments \cite{clerckx2023primer}. We assume that the communication between the source and the users is assisted by the HRIS-UAV. {In contrast, the source to target link is assumed to be established only via the direct path, which avoids the four-hop path loss incurred by RIS-assisted sensing considered in \cite{x.song}. Moreover, HRIS-aided communication can achieve a required sum-rate with lower transmit power \cite{nguyen2023fairness,nguyen2022spectral}, enabling more power to be allocated for sensing.} Finally, we consider battery and solar-powered UAV operation to extend endurance in the considered wireless scenario which enables sustained functionality without continuous external power \cite{singh2023performance}. The proposed system model can be applicable to a broad range of scenarios, including dense IoT scenario, emergency communication, smart factories, and industrial automation.

\subsection{Communication Model} 
\subsubsection{Hybrid RIS}
{In the considered system, the HRIS consists of $N$ elements, of which $N_\mathrm{a}$ are active and $N-N_\mathrm{a}$ are passive. 
The active elements amplify the incident signal through active loads, while the passive elements only reflect it. Since active elements require additional power consumption and hardware circuitry, only a small subset of the HRIS elements, i.e., $N_\mathrm{a}\ll N$, is activated to balance performance and complexity \cite{nguyen2022spectral}.
 Let $\mathcal{I}_\mathrm{a}\subset\{1,2,\dots,N\}$ be the index set of the active elements, with $|\mathcal{I}_\mathrm{a}|=N_\mathrm{a}$. The reflection coefficient of the $n$-th HRIS element is denoted by $\eta_n=|\eta_n|e^{j\theta_n}$, where $\theta_n\in[0,2\pi)$ is the phase shift and $|\eta_n|$ is the amplitude. For passive elements, i.e., $n\notin\mathcal{I}_\mathrm{a}$, the amplitude is constrained by $|\eta_n|\leq1$, since these elements do not introduce signal amplification. In contrast, for active elements, i.e., $n\in\mathcal{I}_\mathrm{a}$, both the phase and amplitude can be controlled, and the amplitude is bounded by $|\eta_n|\leq\eta_{\max}$, where $\eta_{\max}$ denotes the maximum gain provided by the active load \cite{zhang2022active,nguyen2022spectral}. Accordingly, the HRIS coefficient matrix is defined as $\boldsymbol{\Theta}=\operatorname{diag}\{\eta_1,\eta_2,\dots,\eta_N\}$. Specifically, it is decomposed as $\boldsymbol{\Theta}=\boldsymbol{\Phi}+\boldsymbol{\Psi}$, where $\boldsymbol{\Phi}=(\mathbf{I}_N-\mathbf{M}_\mathrm{a})\circ\boldsymbol{\Theta}$ and $\boldsymbol{\Psi}=\mathbf{M}_\mathrm{a}\circ\boldsymbol{\Theta}$ represent the passive and active coefficient matrices, respectively. Here, $\mathbf{M}_\mathrm{a}$ is an $N\times N$ diagonal selection matrix whose diagonal entries are one for the active elements indexed by $\mathcal{I}_\mathrm{a}$ and zero otherwise. Thus, $\mathbf{I}_N-\mathbf{M}_\mathrm{a}$ selects the passive elements, while $\mathbf{M}_\mathrm{a}$ selects the active elements. As a result, $\boldsymbol{\Phi}$ contains only the coefficients of the passive reflecting elements, whereas $\boldsymbol{\Psi}$ contains only the coefficients of the active amplifying elements \cite{nguyen2022spectral}.}

\subsubsection{Channel Model}\label{channle_model} The effective channels from the source to the HRIS-UAV and from the HRIS-UAV to the $k$-th user are given by 
$\tilde{\mathbf{H}}_{\mathrm{sr}} = \beta_{\mathrm{sr}}^{1/2} \mathbf{H}_{\mathrm{sr}} \in \mathbb{C}^{N \times N_\mathrm{t}}$ 
and 
$\tilde{\mathbf{h}}_{\mathrm{r}k} = \beta_{\mathrm{r}k}^{1/2} \mathbf{h}_{\mathrm{r}k} \in \mathbb{C}^{N \times 1}$, respectively. 
Here, $\beta_{\mathrm{x}} = d_{\mathrm{x}}^{-\epsilon_{\mathrm{x}}}$ is the large-scale fading coefficient, where $\mathrm{x} \in \{\mathrm{sr},\mathrm{r}k\}$, $d_{\mathrm{x}}$ denotes the Euclidean distance between the corresponding terminals, and $\epsilon_{\mathrm{x}}$ is the path loss exponent. {Hence, the UAV geometry affects the link quality through the link distance and elevation angle. Specifically, $\epsilon_{\mathrm{x}}$ depends on the elevation angle $\vartheta_{\mathrm{x}}$ of the HRIS-UAV link and is modeled as $\epsilon_{\mathrm{x}}(\vartheta_{\mathrm{x}})=\mathcal{L}(\vartheta_{\mathrm{x}})a_{\mathrm{x}}+b_{\mathrm{x}}$, where $\mathcal{L}(\vartheta_{\mathrm{x}})=\left(1+c_{\mathrm{x}}\exp\left(-g_{\mathrm{x}}(\vartheta_{\mathrm{x}}-c_{\mathrm{x}}\right)\right)^{-1}$ is the LoS probability and $a_{\mathrm{x}}$, $b_{\mathrm{x}}$, $c_{\mathrm{x}}$, and $g_{\mathrm{x}}$ are environment-dependent coefficients~\cite{Solanki_TVT}.}
The small-scale fading matrix from the source to the HRIS-UAV is given as 
$\mathbf{H}_{\mathrm{sr}} = [ \mathbf{h}_{\mathrm{sr},1}, \mathbf{h}_{\mathrm{sr},2}, \dots, \mathbf{h}_{\mathrm{sr},N_\mathrm{t}} ]$, where each 
$\mathbf{h}_{\mathrm{sr},i} = [ h_{\mathrm{sr}1,i}, \dots, h_{\mathrm{sr}N,i} ]^\H$ corresponds to the channel between 
the $i$-th transmit antenna and HRIS-UAV. Similarly, the small-scale channel fading coefficients between the HRIS-UAV and $k$-th user is modeled as $\mathbf{h}_{\mathrm{r}k} = [ h_{\mathrm{r}k,1}, h_{\mathrm{r}k,2}, \dots, h_{\mathrm{r}k,N} ]^\H$. We assume that both the HRIS elements and transmit antennas are placed at half-wavelength spacing \cite{nguyen2023fairness}. Based on this, we consider an independent
and identically distributed (i.i.d.)  Nakagami-$m$ fading model for the channel coefficients, such that $|h_{\mathrm{sr}n,i}| \sim \mathrm{Nak}(m_{\mathrm{sr}}, \Omega_{\mathrm{sr}})$ and $|h_{\mathrm{r}k,n}| \sim \mathrm{Nak}(m_{\mathrm{r}k}, \Omega_{\mathrm{r}k})$, $\forall n, i, k$.

\subsubsection{Signal Model}\label{signal_model}
According to the RSMA protocol, the transmitted stream from the source is given as $\mathbf{s}[l] = \begin{bmatrix} s_{\mathrm{c}}[l], & s_{\mathrm{p},1}[l], & \dots, & s_{\mathrm{p},K}[l], & s_{\mathrm{s}}[l] \end{bmatrix}^{\mathrm{T}}, \quad l=1,\dots,L,$ where $L$ is the length of the communication/radar frame. Here, $s_{\mathrm{c}}[l]$ is the common stream, $s_{\mathrm{p},k}[l]$ is the private stream of user $k$, and $s_{\mathrm{s}}[l]$ is the radar-signal-sensing stream. The corresponding transmitted signal from the source is given as
\begin{align}
    \mathbf{x}[l] = \hat{\mathbf{w}}_{\mathrm{c}} \, s_{\mathrm{c}}[l] + \sum_{j=1}^{K} \hat{\mathbf{w}}_{\mathrm{p},j} \, s_{\mathrm{p},j}[l] + \hat{\mathbf{w}}_{\mathrm{s}} \, s_{\mathrm{s}}[l], \label{Tran_signal}
\end{align}
where $\hat{\mathbf{w}}_{\mathrm{y}} = \sqrt{\alpha_{\mathrm{y}} P_{\mathrm{t}}} \, \mathbf{w}_{\mathrm{y}}, \quad \mathrm{y} \in \{ \mathrm{c}, \mathrm{p}, \mathrm{s} \}.$ Here, $\mathbf{w}_{\mathrm{c}}$ and $\mathbf{w}_{\mathrm{p},k}$ are the beamforming vectors for the common and private streams which are designed using a hybrid zero-forcing and maximum-ratio transmission approach \cite{vu2023performance}, while $\mathbf{w}_{\mathrm{s}}$ is the sensing beamformer. Additionally, $P_{\mathrm{t}}$ denotes the total transmit power and $\alpha_{\mathrm{c}}, \alpha_k,$ and $\alpha_{\mathrm{s}}$ are the power allocation factors for the common, private, and sensing streams, respectively, satisfying  $\alpha_{\mathrm{c}} + \sum_{k=1}^{K} \alpha_k + \alpha_{\mathrm{s}} = 1$. Assuming that the streams $s_\mathrm{c}[l]$, $s_{\mathrm{p},k}[l]$, and $s_\mathrm{s}[l]$ are independent, zero-mean, and unit-power, the covariance matrix of the transmit signal is derived as $\mathbf{R}_x = \mathbb{E}[\mathbf{x}[l] \, \mathbf{x}[l]^\H]=\mathbb{E}[\mathbf{x}_{\mathrm{c}} \mathbf{x}_{\mathrm{c}}^\H] + \sum_{k=1}^{K} \mathbb{E}[\mathbf{x}_{\mathrm{p},k} \mathbf{x}_{\mathrm{p},k}^\H] + \mathbb{E}[\mathbf{x}_{\mathrm{s}} \mathbf{x}_{\mathrm{s}}^\H]$, where each term is given by  $\mathbf{x}_{\mathrm{c}} = \hat{\mathbf{w}}_{\mathrm{c}} \, s_{\mathrm{c}}[l], 
\mathbf{x}_{\mathrm{p},k} = \hat{\mathbf{w}}_{\mathrm{p},k} \, s_{\mathrm{p},k}[l],$ and 
$\mathbf{x}_{\mathrm{s}} = \hat{\mathbf{w}}_{\mathrm{s}} \, s_{\mathrm{s}}[l]$. Since $\mathbb{E}[|s_{\mathrm{y}}[l]|^2] = 1$, we obtained the expression as  $\mathbf{R}_x = \alpha_{\mathrm{c}} P_{\mathrm{t}} \, \mathbf{w}_{\mathrm{c}} \mathbf{w}_{\mathrm{c}}^\H + \sum_{k=1}^{K} \alpha_k P_{\mathrm{t}} \, \mathbf{w}_{\mathrm{p},k} \mathbf{w}_{\mathrm{p},k}^\H + \alpha_{\mathrm{s}} P_{\mathrm{t}} \, \mathbf{w}_{\mathrm{s}} \mathbf{w}_{\mathrm{s}}^\H$ \cite{11048955}. Thus, at the $l$-th time slot, the received signal at the $k$-th communication user is given as 
\begin{align}
y_k[l] &= \sqrt{\beta_\mathrm{sr}\beta_{\mathrm{r}k}} \mathbf{h}_k^\H \big(   \sqrt{\alpha_\mathrm{c} P_\mathrm{t}} \mathbf{w}_\mathrm{c} s_\mathrm{c}[l] + \sum_{j=1}^{K} \sqrt{\alpha_j P_\mathrm{t}}\mathbf{w}_{\mathrm{p},j} s_{\mathrm{p},j}[l] \nonumber \\
&  + \sqrt{\alpha_\mathrm{s} P_\mathrm{t}} \mathbf{w}_\mathrm{s} s_\mathrm{s}[l] \big)   + \sqrt{\beta_{\mathrm{r}k}}  \mathbf{h}_{\mathrm{r}k}^\H \boldsymbol{\Psi }\mathbf{n}_\mathrm{a}+ n_k,
\end{align}
where  $\mathbf{h}_k^\H    =\mathbf{h}^\H_{\mathrm{r}k}\boldsymbol{\Theta}\mathbf {H}_{\mathrm{sr}} \in \mathbb{C}^{1\times N_\mathrm{t}}$  is the cascaded channel from the source to the $k$-th user via the HRIS-UAV, $\mathbf{n}_\mathrm{a}\sim \mathcal{CN}(0,\sigma_\mathrm{a}^2\mathbf{M}_\mathrm{a})$ is the noise due to active elements of HRIS, and $n_k \sim \mathcal{CN}(0, \sigma_k^2)$ is additive white Gaussian noise (AWGN).

\subsubsection{SINR formulation with imperfect CSI and SIC}
We consider that the communication users have iCSI denoted by $\hat{\mathbf{h}}_k $. The relation between the perfect channel state information (pCSI) and iCSI is expressed as
\begin{align}
\label{icsi}
\hat{\mathbf{h}}_k = \xi \mathbf{h}_k + \sqrt{1 - \xi^2} \bar{\mathbf{h}}_k,\end{align} where $\xi \in [0,1]$ denotes the CSI accuracy factor, with $\xi = 1$ corresponding to pCSI \cite{vu2022performance}.
\begin{figure*}[!t]
\begin{align}\label{sinrs}
\gamma_{\mathrm{c},k} &= 
\frac{ \beta_\mathrm{sr}\beta_{\mathrm{r}k}   {\alpha_\mathrm{c} P_\mathrm{t}} | \hat{\mathbf{h}}_k^\H \mathbf{w}_\mathrm{c} |^2 }
{\beta_\mathrm{sr}\beta_{\mathrm{r}k}   {P_\mathrm{t}} \sum_{j=1}^K\alpha_{j} | \hat{\mathbf{h}}_k^\H\mathbf{w}_{\mathrm{p},j} |^2+ \beta_\mathrm{sr}\beta_{\mathrm{r}k}   {P_\mathrm{t}} \alpha_\mathrm{s}\left| \hat{\mathbf{h}}_k^\H \mathbf{w}_\mathrm{s} \right|^2 + \beta_{\mathrm{r}k} \sigma^2_\mathrm{a} V_k + \sigma^2_k }, \\ \label{sinrs_p}
\gamma_{\mathrm{p},k} & = 
\frac{ \beta_\mathrm{sr}\beta_{\mathrm{r}k}   {\alpha_k P_\mathrm{t}} | \hat{\mathbf{h}}_k^\H \mathbf{w}_{\mathrm{p},k} |^2 }
{\beta_\mathrm{sr}\beta_{\mathrm{r}k}   {P_\mathrm{t}} \sum_{j \neq k}\alpha_{j} | \hat{\mathbf{h}}_k^\H\mathbf{w}_{\mathrm{p},j} |^2+ \delta_\mathrm{e}\beta_\mathrm{sr}\beta_{\mathrm{r}k}   {\alpha_\mathrm{c} P_\mathrm{t}} | \hat{\mathbf{h}}_k^\H \mathbf{w}_\mathrm{c} |^2  + \beta_\mathrm{sr}\beta_{\mathrm{r}k}   {P_\mathrm{t}} \alpha_\mathrm{s}\left| \hat{\mathbf{h}}_k^\H \mathbf{w}_\mathrm{s} \right|^2 + \beta_{\mathrm{r}k} \sigma^2_\mathrm{a} V_k + \sigma^2_k }.
\end{align}
\hrule
\end{figure*}Additionally,  $\bar{\mathbf{h}}_k \sim \mathcal{CN}(0, d_k^{-\tau})$, where $d_k$
is the physical distance of the corresponding link and $\tau$ is the path-loss exponent.
Following the single-layer SIC protocol, user $k$
 first decodes $s_{\mathrm{c}}[l]$ while treating the $s_{\mathrm{p},k}[l]$ and $s_{\mathrm{s}}[l]$ as interference and then decodes $s_{\mathrm{p},k}[l]$ \cite{clerckx2023primer}, \cite{vu2023performance}. Thus, the SINRs at the $k$-th user for decoding $s_\mathrm{c}[l]$ and $s_{\mathrm{p},k}[l]$ are, respectively, given in \eqref{sinrs} and \eqref{sinrs_p}, 
where $\delta_\mathrm{e}$ refers to the level of
residual interference arising due to iSIC with $\delta_\mathrm{e}=0$ 
corresponding to the pSIC case and $V_k \triangleq  \sum_{n \in \mathcal{I}_\mathrm{a}} |\eta_n|^2 | {h}_{\mathrm{r}k,n} |^2$. {To manage multiuser interference while preserving the desired signal gain, we adopt a hybrid ZF-MRT precoding structure for the RSMA streams. We aim to design the private precoder to suppress inter-user interference, since each private stream is intended only for its corresponding user. In contrast, the common stream is decoded by all users and, therefore, its precoder should provide nonzero effective gain at all users rather than nulling the signal toward unintended users. Based on this, let us define the estimated channel matrix as $\mathbf{H} = [\hat{\mathbf{h}}_1, \dots, \hat{\mathbf{h}}_K]$ and the diagonal weighting matrix as $\mathbf{F} = \operatorname{diag}(\|\hat{\mathbf{h}}_1\|, \dots, \|\hat{\mathbf{h}}_K\|)$. The private precoding matrix is designed as $\mathbf{W} = \mathbf{H}(\mathbf{H}^{\H}\mathbf{H})^{-1}\mathbf{F}$, which ensures $\hat{\mathbf{h}}_k^{\H}\mathbf{w}_{\mathrm{p},j}=0$ for $j\neq k$, thereby cancelling inter-user interference among the private streams. Moreover, the weighting matrix $\mathbf{F}$ gives
$\hat{\mathbf{h}}_k^{\H}\mathbf{w}_{\mathrm{p},k}
=\|\hat{\mathbf{h}}_k\|$, which provides the same desired
effective-channel gain as MRT for user $k$. Thus, the private precoder for user $k$ is $\mathbf{w}_{\mathrm{p},k}=\mathbf{H}(\mathbf{H}^{\H}\mathbf{H})^{-1}\mathbf{f}_k$, where $\mathbf{f}_k$ is the $k$-th column of $\mathbf{F}$. Since the common stream must be decoded by all users, the common precoder is constructed as $\mathbf{w}_{\mathrm{c}}=\mathbf{W}\mathbbm{1}$, where $\mathbbm{1}\in\mathbb{C}^{K\times1}$ denotes the all-ones vector. Hence, $\mathbf{w}_{\mathrm{c}}=\sum_{k=1}^{K}\mathbf{w}_{\mathrm{p},k}$ and $\hat{\mathbf{h}}_k^{\H}\mathbf{w}_{\mathrm{c}}=\|\hat{\mathbf{h}}_k\|$, which allows the common stream to provide useful effective gain at all users \cite{vu2023performance}.} Based on this design, the SINRs given in \eqref{sinrs} and \eqref{sinrs_p}, for the common and private messages, respectively, can be simplified as
\begin{align}
\gamma_{\mathrm{c},k} = \frac{ \alpha_\mathrm{c} \rho_\mathrm{t} \beta_k U_k }{ \alpha_k \rho_\mathrm{t} \beta_k U_k + \beta_k \alpha_\mathrm{s} \rho_\mathrm{t}{Z}_k + \tilde{\beta}_k V_k + 1 }, \label{gamma_ck}
\end{align}
\begin{align}
\gamma_{\mathrm{p},k} = \frac{ \alpha_k \beta_k \rho_\mathrm{t} U_k }{ \delta_\mathrm{e} \alpha_\mathrm{c} \beta_k \rho_\mathrm{t} U_k + \beta_k \alpha_\mathrm{s} \rho_\mathrm{t}{Z}_k + \tilde{\beta}_k V_k + 1 }, \label{gamma_pk}
\end{align}
where $\rho_\mathrm{t}\triangleq\frac{P_\mathrm{t}}{\sigma_k^2}$, $\beta_k\triangleq\beta_\mathrm{sr}\beta_{\mathrm{r}k} $, and  $\tilde{\beta}_k\triangleq\beta_{\mathrm{r}k} \frac{\sigma_\mathrm{a}^2}{\sigma_k^2}$.  
Additionally,
$U_k \triangleq \left\| \hat{\mathbf{h}}_k \right\|^2$ and ${Z}_k \triangleq \left| \hat{\mathbf{h}}_k^\H \mathbf{w}_\mathrm{s} \right|^2$.

\subsection{Radar Model}
{For the sensing subsystem, we assume that the target lies in the LoS of the source. The source simultaneously transmits data signals to the communication users and probes the target.}
The $N_\mathrm{r} \times 1$ reflected echo signal received at the source is 
\begin{align}
\mathbf{y}_{\mathrm{s}}[l] = \mathbf{D}_{\mathrm{st}}(\theta)\mathbf{x}[l] + \mathbf{n}_{\mathrm{s}}[l] = \beta_\mathrm{s}\mathbf{b}(\theta) \mathbf{a}^\H(\theta) \mathbf{x}[l] + \mathbf{n}_{\mathrm{s}}[l], \label{received_sen}
\end{align}
where $ \beta_\mathrm{s}$ is the complex reflection coefficient correlated with the radar cross-section (RCS) of the target and $ \mathbf{n}_{\mathrm{s}}[l] \sim \mathcal{CN}(0, \sigma_\mathrm{s}^2 \mathbf{I}_{N_\mathrm{r}}) $ is the AWGN noise with variance $ \sigma_\mathrm{s}^2 $. Here,  $\mathbf{D}_{\mathrm{st}}\in \mathbb{C}^{N_\mathrm{r} \times {N_\mathrm{t}}} $ is the overall channel from the source to the target and from the target to the source, $\theta$ is the angle of the target, and $\mathbf{a}(\theta) $ and $\mathbf{b}(\theta)$ are the transmit and receive steering vectors at the source, respectively.
We consider the center of the uniform linear array (ULA) as the reference point and adopt half-wavelength spacing between adjacent transmit/receive antenna elements. Thus, 
the transmit steering vector is modeled as   $
	\mathbf{a}(\theta) = 
	\begin{bmatrix}
		e^{j \pi m \sin(\theta)}
	\end{bmatrix}_{m = -\frac{N_\mathrm{t}-1}{2}}^{\frac{N_\mathrm{t}-1}{2}}
	\in \mathbb{C}^{N_\mathrm{t}\times 1}
$ \cite{steering}. Furthermore, based on \eqref{received_sen}, given arbitrary communication precoders $ \mathbf{w}_\mathrm{c} $ and $ \mathbf{w}_{\mathrm{p},k} $, the optimal sensing beamformer is $\mathbf{w}_\mathrm{s} = \mathbf{a}(\theta)  $, assuming that the target angle is known \cite{nguyen2025performance}. For notational simplicity, we omit the argument 
$\theta$ and denote the transmit steering vector as $\mathbf{a}$ in the subsequent analyses.

\section{Communication Performance Analysis}\label{performance}
{In this section, we present the analytical expressions for the OP, throughput, and ER of the communication users. Specifically, the OP measures communication reliability, while the ER characterizes the average spectral efficiency over the channel fading realizations. The throughput further quantifies the achievable data transmission rate under delay-limited and delay-tolerant transmission modes.} Furthermore, we provide an asymptotic analysis of the OP and ER to gain key insights into system behavior at high SNRs.

\subsection{Outage Probability}
An outage event is said to occur when the received SINR of a user either for the common or private stream falls below a predefined threshold. 
Accordingly, the OP of the $k$-th user is defined as
\begin{align}
\mathcal{P}_{\mathrm{o}_k} &= 1 - \mathbb{P} \left( \gamma_{\mathrm{c},k} > \gamma_{\mathrm{c}}^{\mathrm{th}}, \ \gamma_{\mathrm{p},k} > \gamma_{\mathrm{p},k}^{\mathrm{th}} \right),
\label{out_1}
\end{align}
{where $\gamma_{\mathrm{c}}^{\mathrm{th}}=2^{R^{\mathrm{c}}}-1$ and $\gamma_{\mathrm{p},k}^{\mathrm{th}}=2^{R^{\mathrm{p},k}}-1$ are the predefined threshold values for the common and private SINRs of the $k$-th user, respectively. Here, $R^{\mathrm{c}}$ and $R^{\mathrm{p},k}$ denote the corresponding target rates.} By substituting the expressions for $\gamma_{\mathrm{c},k}$ and $\gamma_{\mathrm{p},k}$ from equations~\eqref{gamma_ck} and~\eqref{gamma_pk} into equation~\eqref{out_1}, and performing algebraic simplifications, we obtain
\begin{align}
\mathcal{P}_{\mathrm{o}_k}  = & 1 - \mathbb{P}\left( \frac{\rho_\mathrm{t} \beta_k U_k}{\beta_k \alpha_\mathrm{s} \rho_\mathrm{t}Z_k + \tilde{\beta}_k V_k + 1} > \frac{\gamma_{\mathrm{c}}^{\mathrm{th}}}{(\alpha_\mathrm{c} - \alpha_k \gamma_{\mathrm{c}}^{\mathrm{th}})} \right. \nonumber \\
&  \left. , \ \frac{\rho_\mathrm{t} \beta_k U_k}{\beta_k \alpha_\mathrm{s} \rho_\mathrm{t}Z_k + \tilde{\beta}_k V_k + 1} > \frac{\gamma_{\mathrm{p},k}^{\mathrm{th}}}{(\alpha_k - \gamma_{\mathrm{p},k}^{\mathrm{th}} (\delta_\mathrm{e}\alpha_\mathrm{c})  )} \right). 
\label{out_2}
\end{align}
Based on max statistics, \eqref{out_2} can be simplified as
\begin{align}
\mathcal{P}_{\mathrm{o}_k}  &= 1 - \mathbb{P}\left( \frac{\rho_\mathrm{t} \beta_k U_k}{\beta_k \alpha_\mathrm{s}\rho_\mathrm{t}Z_k + \tilde{\beta}_k V_k + 1} > \gamma_k^{\mathrm{th}} \right) \nonumber \\
&= \mathbb{P}\left( \frac{\mathcal{X}_{1,k}}{\mathcal{X}_{2,k} + \mathcal{X}_{3,k} + 1} \leq \gamma_k^{\mathrm{th}} \right), \label{out_3}
\end{align}
 where $\mathcal{X}_{1,k}\triangleq \rho_\mathrm{t} \beta_k U_k$ , $ 
\mathcal{X}_{2,k} \triangleq \beta_k \alpha_\mathrm{s} \rho_\mathrm{t}Z_k$, $
\mathcal{X}_{3,k}\triangleq \tilde{\beta}_k V_k$, and $\gamma_k^{\mathrm{th}}\triangleq\max\left( A_{1,k}, A_{2,k} \right)$. Here, 
\begin{align}
  A_{1,k}\triangleq \frac{\gamma_{\mathrm{c}}^{\mathrm{th}}}{\left( \alpha_\mathrm{c} - \alpha_k \gamma_{\mathrm{c}}^{\mathrm{th}} \right)} \, \text{and}   \, A_{2,k} \triangleq \frac{\gamma_{\mathrm{p},k}^{\mathrm{th}}}{\alpha_k - \gamma_{\mathrm{p},k}^{\mathrm{th}} (\delta_\mathrm{e} \alpha_\mathrm{c})}.
\end{align}

\begin{remark}\label{threshold_cond}
We note that \eqref{out_3} is valid only when $\gamma_{\mathrm{c}}^{\mathrm{th}} < \frac{\alpha_\mathrm{c}}{\alpha_k}$ and $\gamma_{\mathrm{p},k}^{\mathrm{th}} < \frac{\alpha_k}{\delta_\mathrm{e} \alpha_\mathrm{c}}$, otherwise, the $k$-th user experiences outage, and the OP attains a ceiling value of one.  
\end{remark}
To facilitate the derivation of the closed form expression for the OP given in \eqref{out_3}, we first present the statistical distribution of the random variables (RVs) $\mathcal{X}_{1,k}$, $\mathcal{X}_{2,k}$, and $\mathcal{X}_{3,k}$ in the following lemma.
\begin{lemma}
\label{lemma:RV}
 The statistical characteristics of the RVs $\mathcal{X}_{1,k}$, $\mathcal{X}_{2,k}$, and $\mathcal{X}_{3,k}$ are 
\begin{itemize}
    \item $\mathcal{X}_{1,k} \sim \mathrm{Gamma} \left( \kappa_{\|\hat{\mathbf{h}}_{k}\|^2}, \frac{\lambda_{\|\hat{\mathbf{h}}_{k}\|^2}}{\rho_\mathrm{t} \beta_k} \right)$ 
with the shape and scale parameters, respectively, given by 
\begin{align}
\kappa_{\|\hat{\mathbf{h}}_{k}\|^2} = \dfrac{\mathbb{E} \left[ \|\hat{\mathbf{h}}_{k}\|^2\right]^2}{\mathrm{Var}\left[\|\hat{\mathbf{h}}_{k}\|^2\right]}, 
\lambda_{\|\hat{\mathbf{h}}_{k}\|^2} =\dfrac{\mathbb{E}\left[\|\hat{\mathbf{h}}_{k}\|^2\right]}{\mathrm{Var}\left[\|\hat{\mathbf{h}}_{k}\|^2\right]},
\label{hk_bold1}
\end{align}
where $\mathrm{Var}\left[\|\hat{\mathbf{h}}_{k}\|^2\right]={\mathbb{E}\left[\|\hat{\mathbf{h}}_{k}\|^4\right]-\mathbb{E}\left[\|\hat{\mathbf{h}}_{k}\|^2\right]^2}$ and 
\begin{align}
\label{hk_bold}
\mathbb{E}[\|\hat{\mathbf{h}}_{k}\|^2] &= \dfrac{\xi^2 N_\mathrm{t} \kappa_{|{h}_{k,i}|^2}^2}{\lambda_{|{h}_{k,i}|^2}} + (1 - \xi^2) N_\mathrm{t} d_k^{-\tau}, \\
\mathbb{E}[\|\hat{\mathbf{h}}_{k}\|^4] &= \dfrac{\kappa_{|{h}_{k,i}|^2} (\lambda_{|{h}_{k,i}|^2}(N_\mathrm{t} + 1) + \kappa_{|{h}_{k,i}|^2})}{\lambda_{|{h}_{k,i}|^2}^2}  \nonumber \\   & \quad+  (1 - \xi^2)^2 N_\mathrm{t}(N_\mathrm{t} + 1) d_k^{-2\tau} \nonumber \\   & \quad + 2 \xi^2 (1 - \xi^2) \dfrac{N_\mathrm{t} \kappa_{|{h}_{k,i}|^2}}{\lambda_{|{h}_{k,i}|^2}} d_k^{-\tau}.   
\end{align}
Here, $\kappa_{|{h}_{k,i}|^2}$ and $\lambda_{|{h}_{k,i}|^2}$ are computed similar to \eqref{hk_bold1}. Their corresponding expectations are given in \eqref{hRk,i} and \eqref{hRk,i_sq}, respectively,
\begin{figure*}[!t]
\begin{align}
\label{hRk,i}
\mathbb{E}\left[ {|{h}_{k,i}|^2} \right] &= \sum_{n=1}^{N} |\eta_n|^2 \mathcal{M}^{(2)}_{|{h}_{\mathrm{sr}_{n,i}}|} \mathcal{M}^{(2)}_{|{h}_{\mathrm{r}k,n}|} + 2 \sum_{m=1}^{N} \sum_{n=m+1}^{N} |\eta_m||\eta_n| \mathcal{M}_{|{h}_{\mathrm{sr}_{m,i}}|} \mathcal{M}_{|{h}_{\mathrm{sr}_{n,i}}|} \mathcal{M}_{|{h}_{\mathrm{r}k,m}|} \mathcal{M}_{|{h}_{\mathrm{r}k,n}|}, \\ 
\label{hRk,i_sq}
\mathbb{E} \left[ {|{h}_{k,i}|}^4 \right] 
&= \sum_{n=1}^{N} |\eta_n|^4 \, \mathcal{M}^{(1)}_{|{h}_{\mathrm{sr}_{n,i}}|} \, \mathcal{M}^{(4)}_{|{h}_{\mathrm{r}k,n}|}  + \sum_{\substack{m=1 \\ n \neq m}}^{N} \sum_{n=1}^{N} |\eta_m|^2 |\eta_n|^2 \, \mathcal{M}^{(2)}_{|{h}_{\mathrm{sr}_{m,i}}|} \, \mathcal{M}^{(2)}_{|{h}_{\mathrm{sr}_{n,i}}|} \, \mathcal{M}^{(2)}_{|{h}_{\mathrm{r}k,m}|} \, \mathcal{M}^{(2)}_{|{h}_{\mathrm{r}k,n}|} \notag \\
&\quad + 4 \sum_{\substack{m=1 \\ n \neq m}}^{N} \sum_{n=1}^{N} |\eta_m|^3 |\eta_n| \, \mathcal{M}^{(3)}_{|{h}_{\mathrm{sr}_{m,i}}|} \, \mathcal{M}^{(1)}_{|{h}_{\mathrm{sr}_{n,i}}|} \, \mathcal{M}^{(2)}_{|{h}_{\mathrm{r}k,m}|} \, \mathcal{M}^{(1)}_{|{h}_{\mathrm{r}k,n}|} \nonumber \\
&\quad + 4 \sum_{\substack{n=1 \\ n \ne l}}^N \sum_{\substack{m=1 \\ m \ne l}}^N \sum_{\substack{l=1 \\ l \ne m, l \ne n}}^N 
|\eta_m| |\eta_n| |\eta_l|^2 
\mathcal{M}^{(2)}_{|{h}_{\mathrm{sr}_{m,i}}|} 
\mathcal{M}^{(1)}_{|{h}_{\mathrm{sr}_{m,i}}|} 
\mathcal{M}^{(1)}_{|{h}_{\mathrm{sr}_{n,i}}|} 
\mathcal{M}^{(2)}_{|{h}_{\mathrm{r}k,l}|} 
\mathcal{M}^{(2)}_{|{h}_{\mathrm{r}k,m}|} 
\mathcal{M}^{(2)}_{|{h}_{\mathrm{r}k,n}|} \notag \\
&\quad + 4 \sum_{\substack{j=1 \\ j \ne l}}^N \sum_{\substack{m=1 \\ m \ne j}}^N \sum_{\substack{n = m+1 \\ n \ne j}}^N 
|\eta_j|^2 |\eta_m| |\eta_n| 
\mathcal{M}^{(2)}_{|{h}_{\mathrm{sr}_{m,j}}|} 
\mathcal{M}^{(2)}_{|{h}_{\mathrm{r}k,j}|} 
\mathcal{M}^{(2)}_{|{h}_{\mathrm{sr}_{m,i}}|} 
\mathcal{M}^{(2)}_{|{h}_{\mathrm{r}k,m}|} 
\mathcal{M}^{(2)}_{|{h}_{\mathrm{sr}_{n,i}}|} 
\mathcal{M}^{(2)}_{|{h}_{\mathrm{r}k,n}|} \nonumber  \\
&\quad + 8 \sum_{\substack{l=1 \\ l \ne m, l \ne n}}^N \sum_{\substack{m=1 \\ m \ne l}}^N \sum_{\substack{n=m+1 \\ n \ne l}}^N 
|\eta_l|^2 |\eta_m| |\eta_n| 
\mathcal{M}^{(2)}_{|{h}_{\mathrm{sr}_{m,l}}|} 
\mathcal{M}^{(2)}_{|{h}_{\mathrm{r}k,l}|} 
\mathcal{M}^{(1)}_{|{h}_{\mathrm{sr}_{m,i}}|} 
\mathcal{M}^{(1)}_{|{h}_{\mathrm{r}k,m}|} 
\mathcal{M}^{(1)}_{|{h}_{\mathrm{sr}_{n,i}}|} 
\mathcal{M}^{(1)}_{|{h}_{\mathrm{r}k,n}|} \notag \\
&\quad + 4 \sum_{\substack{j=1}}^N \sum_{\substack{l=1 \\ l \ne j}}^N \sum_{\substack{m=1 \\ l \ne j \ne m}}^N \sum_{\substack{n=1 \\ l \ne j \ne m \ne n}}^N 
|\eta_j| |\eta_l| |\eta_m| |\eta_n| 
\mathcal{M}^{(1)}_{|{h}_{\mathrm{sr}_{m,j}}|} 
\mathcal{M}^{(1)}_{|{h}_{\mathrm{r}k,j}|} 
\mathcal{M}^{(1)}_{|{h}_{\mathrm{sr}_{m,l}}|} 
\mathcal{M}^{(1)}_{|{h}_{\mathrm{r}k,l}|} 
\mathcal{M}^{(1)}_{|{h}_{\mathrm{sr}_{m,i}}|} 
\mathcal{M}^{(1)}_{|{h}_{\mathrm{r}k,m}|}  \nonumber \\
& \hspace{12cm} \times \mathcal{M}^{(1)}_{|{h}_{\mathrm{sr}_{n,i}}|} 
\mathcal{M}^{(1)}_{|{h}_{\mathrm{r}k,n}|}.
\end{align}
\hrule
\end{figure*}
and the $p$-th moment of $X$ is given as
\begin{align}
\label{nak_mean}
\mathcal{M}^{(p)}_X = \dfrac{\Gamma(m_X + p/2)}{\Gamma(m_X)} \left( \dfrac{m_X}{\Omega_X} \right)^{-p/2},
\end{align}
for 
$X \in \{|{h}_{\mathrm{sr}_{n,i}}|, |{h}_{\mathrm{r}k,n}| \}, \forall n,i$. 

\item  $ \mathcal{X}_{2,k} \sim \mathrm{Gamma} \left( \kappa_{Z_k}, \frac{\lambda_{Z_k}}{\beta_k \alpha_\mathrm{s} \rho_\mathrm{t}} \right)$, with
\begin{align}
\kappa_{Z_k} = \frac{\mathbb{E}[Z_{k}]^2}{\mathrm{Var}\left[Z_{k}\right]} ,\
    \lambda_{Z_k} =\frac{\mathbb{E}[Z_{k}]}{\mathrm{Var}\left[Z_{k}\right]}.
\end{align}
Here, $\mathrm{Var}\left[Z_{k}\right]={\mathbb{E}[Z_{k}^2]}-{\mathbb{E}[Z_{k}]^2}$ and $\mathbb{E}[Z_{k}]$ and $\mathbb{E}[Z_{k}^2]$ given in \eqref{zk} and \eqref{zk_sq}, respectively, with 
\begin{figure*}
\begin{align}
\label{zk}
    \mathbb{E}[Z_{k}] &= \xi^2 \left( \mathbf{a}^{\H} \mathbf{R}_{\mathbf{h}_k}\mathbf{a} + (\mathbf{m}^{\H} \mathbf{a})^2 \right)
    + (1 - \xi^2) d_k^{-\tau} N_\mathrm{t}, \\ 
\label{zk_sq}
    \mathbb{E}[Z_{k}^2] &= \xi^4 (\mathbf{m}^{\H} \mathbf{a})^4 
    + 4 \xi^2 (\mathbf{m}^{\H}\mathbf{a})^2 
    \left( \xi^2 (\mathbf{a}^{\H} \mathbf{R}_{\mathbf{h}_k}\mathbf{a}) + (1 - \xi^2) d_k^{-\tau} N_\mathrm{t} \right) + 2 \left( \xi^2 (\mathbf{a}^{\H} \mathbf{R}_{\mathbf{h}_k}\mathbf{a}) + (1 - \xi^2) d_k^{-\tau} N_\mathrm{t} \right)^2.
\end{align}\hrule
\end{figure*}
\begin{align}
    \mathbf{m} &= \sum_{n=1}^{N} |\eta_n| \, \mathbf{v}_n \, \mathcal{M}_{|{h}_{\mathrm{r}k,n}|}^{(1)}, \\
\mathbf{R}_{\mathbf{h}_k}&\!=\! \sum_{n=1}^{N} \sum_{m=1}^{N} \!|\eta_n||\eta_m|\mathcal{M}_{|{h}_{\mathrm{r}k,n}|}^{(1)} \mathcal{M}_{|{h}_{\mathrm{r}k,m}|}^{(1)} 
    \mathbf{v}_n \mathbf{v}_m^{\H},
\end{align}
where $\mathbf{v}_n \in \mathbb{C}^{N_\mathrm{t}\times 1}$ is a vector whose $i$-th entry is given by $\mathcal{M}^{(1)}_{|{h}_{\mathrm{sr}_{n,i}}|}, \forall n,i$ and can be computed similar to \eqref{nak_mean}.
\item $   \mathcal{X}_{3,k} \sim \mathrm{Gamma} \left( \kappa_{V_k}, \frac{\lambda_{V_k}}{\tilde{\beta}_k} \right)$,
where
\begin{align}
   \kappa_{V_k} = \frac{\mathbb{E}[V_{k}]^2}{\mathrm{Var}\left[V_k\right]}, \
    \lambda_{V_k} =\frac{\mathbb{E}[V_{k}]}{\mathrm{Var}\left[V_k\right]},
\end{align}
and $\mathrm{Var}\left[V_k\right]={\mathbb{E}[V_{k}^2]}-{\mathbb{E}[V_{k}]^2}$ with
\begin{align}
    \mathbb{E}[V_k] &= \sum_{n \in \mathcal{I}_a} |\eta_n|^2 \, \mathcal{M}^{(2)}_{|{h}_{\mathrm{r}k,n}|}, \\
 \mathbb{E}[V_k^2] &= \sum_{n \in \mathcal{I}_a} |\eta_n|^4 \, \mathcal{M}^{(4)}_{|{h}_{\mathrm{r}k,n}|} + 2 \!\!\sum_{n \in \mathcal{I}_a} \sum_{m \neq n} |\eta_n|^2 \nonumber \\
    &\quad \times |\eta_m|^2  
    \!\mathcal{M}^{(2)}_{|{h}_{\mathrm{r}k,n}|}  \mathcal{M}^{(2)}_{|{h}_{\mathrm{r}k,m}|},
\end{align}
and $\mathcal{M}^{(p)}_{|{h}_{\mathrm{r}k,n}|}$ can be computed using \eqref{nak_mean}.
\end{itemize} 
\begin{IEEEproof}
 	See Appendix~\ref{app:lemma_rv}.	\end{IEEEproof}
\end{lemma}

Next, we use the results from Lemma~\ref{lemma:RV} in \eqref{out_3} to derive an accurate expression for the OP of the $k$-th user in the following theorem.

\begin{thm}
\label{thm:out_accurate}
The expression for the OP of the considered system is accurately given in \eqref{out_accurate}, where
$
x_j = \cos\left( \frac{(2j - 1)\pi}{2T_1} \right)$ with $T_1$ being the Gauss–Chebyshev quadrature (GCQ) accuracy parameter, 
$\gamma^\mathrm{th}_{k}$ is given in \eqref{out_3} and $\beta_k=\beta_\mathrm{sr}\beta_{\mathrm{r}k} $, and  $\tilde{\beta}_k=\beta_{\mathrm{r}k} \frac{\sigma_a^2}{\sigma_k^2}$. Additionally,
$\kappa_X $ and $ \lambda_X$ for $ X \in \{ \| \hat{\mathbf{h}}_{k} \|^2, Z_k, V_{k} \}$ are provided in Lemma~\ref{lemma:RV}.
\begin{figure*}
 \begin{align}
 \label{out_accurate}
\mathcal{P}_{\mathrm{o}_k}  & =
    \frac{2\pi}{{(\gamma_k^{\mathrm{th}})^2} T_1 \Gamma\left(\kappa_{\|\hat{\mathbf{h}}_{k}\|^2}\right)} 
    \sum_{j=1}^{T_1} \frac{\sqrt{1 - x_j^2}}{(1 - x_j)^2}\gamma\left( \kappa_{\|\hat{\mathbf{h}}_{k}\|^2}, 
    \frac{{\lambda_{\|\hat{\mathbf{h}}_{k}\|^2}}}{\rho_\mathrm{t} \beta_k} 
    \left( \frac{1 + x_j}{1 - x_j} + \gamma_{k}^{\mathrm{th}} \right) \right)\left( \frac{\lambda_{Z_k}}{\beta_k  \alpha_{\mathrm{s}} \rho_\mathrm{t}} \right)^{\kappa_{Z_k}}
    \left( \frac{\lambda_{V_k}}{\tilde{\beta}_k} \right)^{\kappa_{V_k}}
   \nonumber \\ 
 & \quad \times  \bar{\mathrm{H}}_{2,2}^{0,2} \left[\exp\left(\frac{1 + x_j}{\left(1 - x_j\right)\gamma_{k}^{\mathrm{th}} } \right)\ \bigg \vert \begin{array}{c}\left( 1 - \frac{\lambda_{Z_k}}{\beta_k \alpha_\mathrm{s} \rho_\mathrm{t}},\, 1,\, \kappa_{Z_k}\right), \left( 1 - \frac{\lambda_{V_k}}{\tilde{\beta}_k}, 1,{\kappa_{V_k}} \right)\\\left( -\frac{\lambda_{Z_k}}{\beta_k \alpha_\mathrm{s} \rho_\mathrm{t}},1, \kappa_{Z_k} \right),\left( -\frac{\lambda_{V_k}}{\tilde{\beta}_k},\, 1,\, {\kappa_{V_k}} \right) \end{array} \right].
\end{align} 
\hrule
\end{figure*} 	\begin{IEEEproof}
 	See Appendix \ref{app:outage}.
 	\end{IEEEproof}	
\end{thm}

\subsection{Asymptotic Outage Probability}
To gain insights into the diversity order and coding gain of the system, we analyze the asymptotic behavior of the OP at high SNRs. The OP at high-SNR regime, is approximated as 
\begin{align}
\mathcal{P}_{\mathrm{o}_k}^{\infty} \approx   \left(\mathcal{G}_k\rho_\mathrm{t}\right)^{-\mathcal{D}_k}, \label{outage_asy_eqn}
\end{align}
where $\mathcal{D}_k$ is the diversity order of the $k$th user, given by
$\mathcal{D}_k$
    \begin{align} 
    \label{diversity_order}\mathcal{D}_k\triangleq {{\kappa_{\|\hat{\mathbf{h}}_{k}\|^2}}+\kappa_{Z_k}}.   
    \end{align}
Here, ${\kappa_{\|\hat{\mathbf{h}}_{k}\|^2}}$, and $\kappa_{Z_k}$ are given in Lemma~\ref{lemma:RV} and
$\mathcal{G}_k$ is the coding gain in \eqref{coding_gain}, where
\begin{figure*}
    \begin{align}
        \label{coding_gain}\mathcal{G}_k & \triangleq \beta_k \Bigg(\frac{2\pi}{ T_1{(\gamma_k^{\mathrm{th}})^2} } 
    \sum_{j=1}^{T_1} \frac{\sqrt{1 - x_j^2}}{(1 - x_j)^2}  \left( \frac{\lambda_{Z_k}}{\alpha_\mathrm{s}} \right)^{\kappa_{Z_k}} 
\left( \frac{\lambda_{V_k}}{\tilde{\beta}_k} \right)^{\kappa_{V_k}}  \times \frac{{\kappa^{-1}_{\|\hat{\mathbf{h}}_{k}\|^2}}}{\Gamma\left(\kappa_{\|\hat{\mathbf{h}}_{k}\|^2}\right)} \left( 
    {{\lambda_{\|\hat{\mathbf{h}}_{k}\|^2}}} 
    \left( \frac{1 + x_j}{1 - x_j} + \gamma_{k}^{\mathrm{th}} \right) \right)^{\kappa_{\|\hat{\mathbf{h}}_{k}\|^2}} \nonumber \\
    &   \hspace{8cm}\times \bar{\mathrm{H}}_{2,2}^{0,2} \left[\exp\left(\frac{1 + x_j}{\left(1 - x_j\right)\gamma_{k}^{\mathrm{th}} } \right)\ \bigg \vert \begin{array}{c}\Delta_{1,k}\\\Delta_{2,k}\end{array} \right] \Bigg)^{\frac{-1}{{\kappa_{\|\hat{\mathbf{h}}_{k}\|^2}}+\kappa_{Z_k}}}.
    \end{align}
    \hrule
    \end{figure*} 
\begin{align}
		\label{delta1}\Delta_{1,k}&\triangleq
		\Big\{\,
		\big(1,\,1,\,\kappa_{Z_k}\big),\;
		\big(1-\tfrac{\lambda_{V_k}}{\tilde{\beta}_k},\,1,\,\kappa_{V_k}\big)
		\,\Big\},
        \end{align}
        \begin{align}
        \label{delta2}
\Delta_{2,k}&\triangleq
		\Big\{\,
		\big(0,\,1,\,\kappa_{Z_k}\big),\;
		\big(-\tfrac{\lambda_{V_k}}{\tilde{\beta}_k},\,1,\,\kappa_{V_k}\big)
		\,\Big\}.
	\end{align}
    \begin{IEEEproof}
        {
As $\rho_\mathrm{t}\to\infty$, the argument of the lower-incomplete Gamma function in \eqref{out_accurate} tends to zero, i.e., $\frac{\lambda_{\|\hat{\mathbf{h}}_k\|^2}}{\rho_\mathrm{t}\beta_k}\big(\frac{1+x_j}{1-x_j}+\gamma_k^{\mathrm{th}}\big)\to 0$. Hence, by using the approximation $\gamma(\alpha,x)\underset{x\to 0}{\simeq}\frac{x^{\alpha}}{\alpha}$, the term $\gamma\big(\kappa_{\|\hat{\mathbf{h}}_k\|^2},\frac{\lambda_{\|\hat{\mathbf{h}}_k\|^2}}{\rho_\mathrm{t}\beta_k}\big(\frac{1+x_j}{1-x_j}+\gamma_k^{\mathrm{th}}\big)\big)$ in \eqref{out_accurate} is approximated as $\frac{1}{\kappa_{\|\hat{\mathbf{h}}_k\|^2}}\bigg(\frac{\lambda_{\|\hat{\mathbf{h}}_k\|^2}}{\beta_k}\big(\frac{1+x_j}{1-x_j}+\gamma_k^{\mathrm{th}}\big)\bigg)^{\kappa_{\|\hat{\mathbf{h}}_k\|^2}}\rho_\mathrm{t}^{-\kappa_{\|\hat{\mathbf{h}}_k\|^2}}$. Thus, this term is proportional to $\rho_\mathrm{t}^{-\kappa_{\|\hat{\mathbf{h}}_k\|^2}}$. Next, in \eqref{out_accurate}, the term $\big(\frac{\lambda_{Z_k}}{\beta_k\alpha_\mathrm{s}\rho_\mathrm{t}}\big)^{\kappa_{Z_k}}$ is proportional to $\rho_\mathrm{t}^{-\kappa_{Z_k}}$, whereas the term $\big(\frac{\lambda_{V_k}}{\tilde{\beta}_k}\big)^{\kappa_{V_k}}$ is independent of $\rho_\mathrm{t}$. Finally, by letting $\rho_\mathrm{t}\to\infty$ in the parameters of the Fox-$\mathrm{H}$ term in \eqref{out_accurate}, i.e., setting $\frac{\lambda_{Z_k}}{\beta_k\alpha_\mathrm{s}\rho_\mathrm{t}}\to 0$, the Fox-$\mathrm{H}$ parameters reduce to the $\rho_\mathrm{t}$-independent sets $\Delta_{1,k}$ and $\Delta_{2,k}$ in \eqref{delta1} and \eqref{delta2}. Thus, the OP in \eqref{out_accurate} can be expressed as $\mathcal{P}_{\mathrm{o}_k}^{\infty}\simeq(\mathcal{G}_k\rho_\mathrm{t})^{-\mathcal{D}_k}$, where $\mathcal{D}_k=\kappa_{\|\hat{\mathbf{h}}_k\|^2}+\kappa_{Z_k}$ and the coding gain $\mathcal{G}_k$ is given in \eqref{coding_gain}.}
    \end{IEEEproof}
\begin{remark}
It is observed from \eqref{diversity_order} along with Lemma~\ref{lemma:RV}, that the diversity order and coding gain depend on key system parameters, namely, (i) the iCSI parameter $\xi$, (ii) $N_\mathrm{t}$ and $N$, and (iii) Nakagami-$m$ fading parameters $m_X$, $X \in \{|{h}_{\mathrm{sr}_{n,i}}|, |{h}_{\mathrm{r}k,n}| \}, \forall n,i$. Furthermore, the diversity order is independent of the SIC error $\delta_\mathrm{e}$ and noise due to the active elements. In contrast, the coding gain $\mathcal{G}_k$ depends on $\delta_\mathrm{e}$ and the active noise.  This suggests that the presence of SIC errors and active noise causes a rightward shift in the OP versus SNR curves, indicating a loss in the coding gain. However, the slope of the OP versus SNR curves remains unaffected by the SIC errors $\delta_\mathrm{e}$ and the active noise of the HRIS.
 \end{remark} 
 
\subsection{Ergodic Rate}
The ER is defined as the long-term average of the instantaneous achievable rate over all the channel realizations. Mathematically, the ER of the considered system is defined as
\begin{align}
\label{rate_1}
    \mathcal{R}_{\mathrm{sum}} = \min_{k \in \{1,2,\ldots,K\}} 
\left( \mathcal{R}_{\mathrm{c},k} \right) + \sum_{k=1}^{K} \mathcal{R}_{\mathrm{p},k},
\end{align}
where $\mathcal{R}_{\mathrm{c},k}$ and $\mathcal{R}_{\mathrm{p},k}$ are the ER of the common and private messages, respectively. We present the accurate expressions for $\mathcal{R}_{\mathrm{c},k}$ and $\mathcal{R}_{\mathrm{p},k}$ in the following, which allows us to analyze the sum rate of the considered system.

\begin{thm}
\label{thm:rate_accurate}
    The unified expressions for the ERs of the common and
private messages are given in \eqref{rate_accurate},
\begin{figure*}[t]
\begin{align}
\label{rate_accurate}
\mathcal{R}_{\mathrm{x}, k} &= \frac{2\pi}{T_2 \ln(2)} \sum_{i=1}^{T_2} \frac{\sqrt{1 - y_i^2}}{1 - y_i^2} \left( \frac{1}{2} - \left( \frac{\nu_1 (1 - y_i)}{1 + y_i} - \nu_2 \right)^2 \right.\notag\\ 
& \quad\left. \times  \frac{\pi}{T_1 \Gamma\left(\kappa_{\|\hat{\mathbf{h}}_{k}\|^2}\right)} \sum_{j=1}^{T_1} \frac{\sqrt{1 - x_j^2}}{(1 - x_j^2)^2} 
\gamma\left(\kappa_{\|\hat{\mathbf{h}}_{k}\|^2}, 
    \frac{{\lambda_{\|\hat{\mathbf{h}}_{k}\|^2}}}{\rho_\mathrm{t} \beta_k}  \left( \frac{1 + x_j}{1 - x_j} + \left(\frac{\nu_1 (1 - y_i)}{1 + y_i} - \nu_2 \right)^{-1} \right) \right) \left( \frac{\lambda_{Z_k}}{\beta_k  \alpha_{\mathrm{s}} \rho_\mathrm{t}} \right)^{\kappa_{Z_k}}
  \right. \notag\\
& \quad \left. \times 
  \left( \frac{\lambda_{V_k}}{\tilde{\beta}_k} \right)^{\kappa_{V_k}}\bar{\mathrm{H}}_{2,2}^{0,2} \left[ \exp \left( \frac{1 + x_j}{1 - x_j}  \left(\frac{\nu_1 (1 - y_i) - \nu_2}{1 + y_i}\right) \right)\middle|
\begin{matrix}
\left( 1 - \frac{\lambda_{Z_k}}{\beta_k \alpha_\mathrm{s} \rho_\mathrm{t}},\, 1,\, \kappa_{Z_k}\right), \left( 1 - \frac{\lambda_{V_k}}{\tilde{\beta}_k}, 1,{\kappa_{V_k}} \right)\\\left( -\frac{\lambda_{Z_k}}{\beta_k \alpha_\mathrm{s} \rho_\mathrm{t}},1, \kappa_{Z_k} \right),\left( -\frac{\lambda_{V_k}}{\tilde{\beta}_k},1,{\kappa_{V_k}} \right)
\end{matrix}
\right] \right).
\end{align}
\hrule
\end{figure*}
where $\mathrm{x} \in \{\mathrm{c}, \mathrm{p} \}$ and 
\begin{align}
\{ \nu_1, \nu_2\} = 
\begin{cases}
\{\alpha_\mathrm{c}, \alpha_k \}, & \text{if } \mathrm{x} = \mathrm{c}, \nonumber  \\
\{\alpha_k, \delta_\mathrm{e}\alpha_\mathrm{c} \}, & \text{if } \mathrm{x} = \mathrm{p}.
\end{cases}
\end{align}
Additionally, 
\begin{align}
x_j = \cos\left( \frac{(2j - 1)\pi}{2T_1} \right) \,\text{and} \,\,y_i = \cos\left( \frac{(2i - 1)\pi}{2T_2} \right),
\end{align}
with $T_1$ and $T_2$ as the GCQ accuracy parameters.
\begin{IEEEproof}
 	See Appendix~\ref{app:rate}.
 	\end{IEEEproof}
\end{thm}
\begin{remark} \label{ER_asy_eqn} At high transmit SNR, the ER for the common message can be simplified using \eqref{sinrs} as 
\begin{align}\mathcal{R}_{\mathrm{c},k}^{\infty} \approx \log_2\left(1 + \frac{\alpha_\mathrm{c} \beta_k U_k}{\alpha_k \beta_k U_k + \alpha_\mathrm{s} \beta_k{Z}_k}\right).\label{Rck_inf}
\end{align}
It can be seen from \eqref{Rck_inf}
that the ER saturates at high SNRs owing to residual sensing interference. Similarly, the asymptotic ER for the private message with iSIC is 
\begin{align}
\mathcal{R}_{\mathrm{p},k,\mathrm{iSIC}}^{\infty} \approx \log_2\left(1 + \frac{\alpha_k \beta_k U_k}{\alpha_\mathrm{c} \delta_\mathrm{e} \beta_k U_k + \alpha_\mathrm{s} \beta_k{Z}_k} \right), \label{Rpk_inf_iSIC}
\end{align}
where the impact of iSIC is captured through the factor $\delta_\mathrm{e}$. In the case of pSIC, i.e., $\delta_\mathrm{e}=0$, the private ER is
\begin{align}
\mathcal{R}_{\mathrm{p},k,\mathrm{pSIC}}^{\infty} \approx \log_2\left(1 + \frac{\alpha_k \beta_k U_k \rho_\mathrm{t}}{\rho_\mathrm{t} \alpha_\mathrm{s} \beta_k{Z}_k} \right). \label{Rpk_inf_pSIC}
\end{align}
It follows from \eqref{Rpk_inf_iSIC} and \eqref{Rpk_inf_pSIC} that for both $\mathcal{R}_{\mathrm{p},k,\mathrm{iSIC}}^{\infty}$ and $\mathcal{R}_{\mathrm{p},k,\mathrm{pSIC}}^{\infty}$ the high-SNR slope approaches zero. This is primarily due to two factors: (i) interference due to the sensing target and (ii) residual interference from iSIC.  
In the absence of a sensing target, i.e., in a conventional communication-only system and under pSIC, the asymptotic ER for the private message can be expressed as
$\tilde{\mathcal{R}}_{\mathrm{p},k,\mathrm{pSIC}}^{\infty} \approx \log_2\left(1+{\alpha_k \beta_k U_k \rho_\mathrm{t}}\right)$. This
highlights that the $\tilde{\mathcal{R}}_{\mathrm{p},k,\mathrm{pSIC}}^{\infty}$ depends on the path loss $\beta_k$, power coefficient $\alpha_k$, and increases with SNR.
\end{remark}

\subsection{Throughput}
In wireless systems, the throughput quantifies the effective data rate achieved under varying channel conditions. It is typically evaluated under two modes: delay-limited (DL) and delay-tolerant (DT) transmissions.
\subsubsection{Delay-Limited Transmission Mode}
In this mode, the source transmits at a constant target rate, regardless of the channel variations. However, due to fading, transmission outages may occur \cite{parihar2025performance}. The system throughput is given by
\begin{align} \mathcal{T}_{\mathrm{DL}}=\left(1-\mathcal{P}_{\mathrm{o}_k}^{j} |_{A_{1,k}} \right) R^{\mathrm{c}}+\left(1-\mathcal{P}_{\mathrm{o}_k}^{j} \right) R^{\mathrm{p},k},\label{DL_eqn} 
\end{align}
where $\mathcal{P}_{\mathrm{o}_k}^{j} \big|_{A_{1,k}}$ and $\mathcal{P}_{\mathrm{o}_k}^{j}$ denote the OP corresponding to the common and private messages of the $k$-th user, respectively. If $R^{\mathrm{c}} = R^{\mathrm{p},k} = R$, the maximum achievable throughput is $2R$ under ideal conditions with a negligible outage.
\subsubsection{Delay-Tolerant Transmission Mode}
In the DT mode, the source transmits at a fixed rate up to the ER \cite{singh2023performance}. Thus, the system throughput is given by
\begin{align}
\mathcal{T}_{\mathrm{DT}}&=\mathcal{R}_{\mathrm{sum}}, \label{DT_eqn}
\end{align}
where $\mathcal{R}_{\mathrm{sum}}$ denotes the ergodic sum rate across all users. This mode generally achieves a higher throughput than the DL mode because it leverages the time-averaged channel behavior.
\begin{remark}
The throughput expressions in \eqref{DL_eqn} and \eqref{DT_eqn} highlight the tradeoff between reliability and latency. In the DL mode, the throughput is constrained by the OP, making system design parameters, such as the RIS configuration, UAV placement, and RSMA power allocation, critical for minimizing outages and improving performance. In contrast, the DT mode mitigates the impact of outages by averaging fading, leading to a higher achievable throughput. This underscores the advantage of operating under relaxed latency constraints when prioritizing long-term performance.
\end{remark}

\section{Sensing Performance Analysis}\label{performance_Sensing}
{In this section, we derive the expressions for sensing metrics namely, PoD, PoFA, and CRLB. The PoD and PoFA characterize target detection performance, while the CRLB quantifies the lower bound on the target angle estimation accuracy.} 

\subsection{Probability of Detection }
To assess the radar target detection capability, we formulated a binary hypothesis test based on the received echo signal \cite{ghosh2024performance}. The source applies receive beamforming with a beamforming vector $\mathbf{u}$, resulting in the scalar test statistic as
\begin{align}
    \tilde{y}[l] = \mathbf{u}^\H \mathbf{y}_s[l].
\end{align}
From \eqref{received_sen}, the hypotheses are explicitly defined as
\begin{align}
\label{hypotheses}
    \tilde{y}[l] =
\begin{cases}
	\mathbf{u}^\H \left(  \mathbf{n}_{\mathrm{s}}[l] \right), & \mathcal{H}_0~(\text{target absent}) \\
	\mathbf{u}^\H \left(\beta_\mathrm{s}\mathbf{b}\mathbf{a}^\H \mathbf{x}[l] + \mathbf{n}_{\mathrm{s}}[l] \right), & \mathcal{H}_1~(\text{target present})
\end{cases}
\end{align}
where, the beamformer 
$\mathbf{u}$ is designed to maximize the radar SINR, $\mathbf{R}_{\mathrm{s}}=\alpha_\mathrm{s} P_\mathrm{t}\,\mathbf{w}_\mathrm{s}\mathbf{w}_\mathrm{s}^\H$, $\mathbf{R}_{\tilde{\mathbf{x}}}=\alpha_\mathrm{c} P_\mathrm{t}\,\mathbf{w}_\mathrm{c}\mathbf{w}_\mathrm{c}^\H+\sum_{k=1}^{K}\alpha_k P_\mathrm{t}\,\mathbf{w}_{\mathrm{p},k}\mathbf{w}_{\mathrm{p},k}^\H$, and $\mathbf{R}_{\mathrm{x}}=\mathbf{R}_{\mathrm{s}}+\mathbf{R}_{\tilde{\mathbf{x}}}.$ Based on \eqref{hypotheses}, the resulting radar SINR can be expressed as
\begin{align}
\tilde{\gamma}_r=\frac{| \beta_\mathrm{s}|^2\,\mathbf{u}^\H \mathbf{D}_{\mathrm{st}} \mathbf{R}_{\mathrm{s}} \mathbf{D}_{\mathrm{st}}^\H \mathbf{u}}		{\sigma_\mathrm{s}^2\,\mathbf{u}^\H\mathbf{u}+| \beta_\mathrm{s}|^2\,\mathbf{u}^\H\mathbf{D}_{\mathrm{st}}\mathbf{R}_{\tilde{\mathbf{x}}}\mathbf{D}_{\mathrm{st}}^\H\mathbf{u}}.\label{gamma_r_tilda}
\end{align}
We note that the radar SINR is a generalized Rayleigh quotient optimization problem \cite{fishler2006spatial}, thus the receive beamformer can be modeled as
$\mathbf{u} = \left( \sigma_\mathrm{s}^2 \mathbf{I} + | \beta_\mathrm{s}|^2\,\mathbf{D}_{\mathrm{st}} \mathbf{R}_{\tilde{\mathbf{x}} }\mathbf{D}_{\mathrm{st}}^\H \right)^{-1} \mathbf{D}_{\mathrm{st}} \mathbf{w}_{\mathrm{s}}$.
Furthermore, from \eqref{hypotheses}, under both hypotheses, the beamformed output $\tilde{y}[l]$ is distributed as $\tilde{y}[l]|\mathcal{H}_0\sim\mathcal{CN}(0,\Psi_0)$ and $\tilde{y}[l]|\mathcal{H}_1\sim\mathcal{CN}(0,\Psi_1)$, where
\begin{align}
\Psi_0 & \triangleq \sigma_\mathrm{s}^2 \,\mathbf{u}^\H\mathbf{u}, \\
\Psi_1 & \triangleq |\beta_\mathrm{s}|^2\,\mathbf{u}^\H \mathbf{D}_{\mathrm{st}} \mathbf{R}_{\mathrm{x}} \mathbf{D}_{\mathrm{st}}^\H \mathbf{u} + \sigma_\mathrm{s}^2 \,\mathbf{u}^\H\mathbf{u}.
\end{align}
By applying the Neyman-Pearson criterion for optimal detection, the detection rule is expressed using the test statistic $Y=|\tilde{y}[l]|^2$ as
  $  Y \underset{\mathcal{H}_0}{\overset{\mathcal{H}_1}{\gtrless}} \gamma_\mathrm{s}^{\text{th}}$,
where $\gamma_\mathrm{s}^{\text{th}}$ is  predetermined detection threshold. Thus, the distribution of $Y$ can be expressed as
\begin{align}
Y \sim 
\begin{cases}
\frac{\Psi_0}{2} \chi^2_2, \mathcal{H}_0,\\
\frac{\Psi_1}{2} \chi^2_2,\mathcal{H}_1,
\end{cases}   
\end{align}
where $\chi^2_2$ is the central chi-square distribution with two degrees of freedom. Based on this, the PoFA can be expressed as follows
\begin{align}\label{eq:Pfa_def}
P_{\mathrm{FA}} =\mathbb{P}(Y>\gamma_{\mathrm{s}}^{\text{th}}|\mathcal{H}_0)=1-F_{\chi_2^2}\left(\frac{2\gamma_{\mathrm{s}}^{\text{th}}}{\Psi_0}\right).
\end{align}
Similarly, the expression for the PoD can be mathematically given as
\begin{align}
\label{detection}
P_{\mathrm{D}} = \mathbb{P}(Y > \gamma_{\mathrm{s}}^{\text{th}}|\mathcal{H}_1)
= 1-F_{\chi^2_2}\left(\frac{2\gamma_{\mathrm{s}}^{\text{th}}}{\Psi_1}\right),
\end{align}
where $F_{\chi^2_2}(x)=\gamma\left(1,\frac{x}{2}\right)$ is the cumulative distribution function (CDF) of $\chi^2_2$.

\begin{remark}
From \eqref{eq:Pfa_def} and \eqref{detection}, it can be inferred that $P_{\mathrm{FA}}$ depends only on $\gamma_{\mathrm{s}}^{\mathrm{th}}/\Psi_0$ and remains fixed for given noise power. In contrast, $P_{\mathrm{D}}$ improves with larger $N_\mathrm{t}$ or $\alpha_\mathrm{s}$, since these parameters increase  $\Psi_1$, thereby reducing $\gamma_{\mathrm{s}}^{\mathrm{th}}/\Psi_1$.
\end{remark}

\subsection{Cramér-Rao Lower Bound}
To analyze the fundamental performance bound for angle estimation, we derive the CRLB for target angle $\theta$. The received echo signal in \eqref{received_sen} can be rewritten as
\begin{align}
    \mathbf{y}_{\mathrm{s}}[l] = \mathbf{r}[l] + \mathbf{n}_{\mathrm{s}}[l],
\end{align}
where $\mathbf{r}[l] = \beta_\mathrm{s}\mathbf{D}_{\mathrm{st}}\mathbf{x}[l]$.
Let the unknown parameter vector be defined as $\boldsymbol{\omega} = [\theta, \boldsymbol{\beta}_\mathrm{s}]^\T$ with $\boldsymbol{\beta}_\mathrm{s}=[\mathrm{Re}\{\beta_\mathrm{s}\}, \mathrm{Im}\{\beta_\mathrm{s}\}]$. The Fisher information matrix (FIM), denoted by $\mathbf{F} \in \mathbb{R}^{3 \times 3}$, associated with the estimation of $\boldsymbol{\omega}$ is given by
\begin{align}
\label{fim}
\mathbf{F} =
\begin{bmatrix}
\mathbf{F}_{\theta \theta} & \mathbf{F}_{\theta \boldsymbol{\beta}_\mathrm{s}} \\
\mathbf{F}_{\theta \boldsymbol{\beta}_\mathrm{s}}^{\T} & \mathbf{F}_{\boldsymbol{\beta}_\mathrm{s} \boldsymbol{\beta}_\mathrm{s}}
\end{bmatrix}.
\end{align}

The elements of the FIM are computed as
\begin{align}
\left[ \mathbf{F} \right]_{i,j} = \frac{2}{\sigma_\mathrm{s}^2} \, \mathrm{Re} \left\{ \sum_{l=1}^{L} \frac{\partial \mathbf{r}[l]^{\H}}{\partial \omega_i} \frac{\partial \mathbf{r}[l]}{\partial \omega_j} \right\}, \quad i,j \in \{1,2,3\}.
\end{align}

The expressions for the elements of the FIM are 
\begin{align}
\label{f_tt}
\mathbf{F}_{\theta \theta} 
&= \frac{2L|\beta_\mathrm{s}|^2}{\sigma_\mathrm{s}^2} \mathrm{tr} \left( \dot{\mathbf{D}}_{\mathrm{st}} \mathbf{R}_\mathrm{x} \dot{\mathbf{D}}_{\mathrm{st}}^{\H} \right)  \in \mathbb{R}^{1 \times 1}, \\
\label{f_tb}
\mathbf{F}_{\theta \boldsymbol{\beta}_\mathrm{s}} 
&= \frac{2L}{\sigma_\mathrm{s}^2} \, \mathrm{Re} \left\{ \beta_\mathrm{s}^{*} \, \mathrm{tr} \left( \mathbf{D}_{\mathrm{st}} \mathbf{R}_\mathrm{x} \dot{\mathbf{D}}_{\mathrm{st}}^{\H} \right) [1,j] \right\} \in \mathbb{R}^{1 \times 2}, \\
\mathbf{F}_{\boldsymbol{\beta}_\mathrm{s} \boldsymbol{\beta}_\mathrm{s}} 
&= \frac{2L}{\sigma_\mathrm{s}^2} \, \mathrm{tr} \left( \mathbf{D}_{\mathrm{st}} \mathbf{R}_\mathrm{x} \mathbf{D}_{\mathrm{st}}^{\H} \right) \mathbf{I}_2 \in \mathbb{R}^{2 \times 2},
\end{align}
where the transmit covariance matrix is $\mathbf{R}_\mathrm{x} = \alpha_{\mathrm{c}} P_\mathrm{t} \, \mathbf{w}_{\mathrm{c}} \mathbf{w}_{\mathrm{c}}^\H + \sum_{k=1}^{K} \alpha_k P_\mathrm{t} \, \mathbf{w}_{\mathrm{p},k} \mathbf{w}_{\mathrm{p},k}^\H + \alpha_{\mathrm{s}} P_\mathrm{t} \, \mathbf{w}_{\mathrm{s}} \mathbf{w}_{\mathrm{s}}^\H,$
 with $\mathbf{w}_{\mathrm{c}}$, $\mathbf{w}_{\mathrm{p},k}$, and $\mathbf{w}_{\mathrm{s}}$ defined in Section~\ref{signal_model}. Additionally, 
\begin{align}
\mathbf{D}_{\mathrm{st}} = \mathbf{b} \mathbf{a}^{\H} \ \text{and} \
\dot{\mathbf{D}}_{\mathrm{st}}= \dot{\mathbf{b}} \mathbf{a}^{\H} + \mathbf{b} \dot{\mathbf{a}}^{\H},
\end{align}
where $\dot{\mathbf{a}} = \frac{\partial \mathbf{a}}{\partial \theta}$ and $\dot{\mathbf{b}} = \frac{\partial \mathbf{b}}{\partial \theta}$. Using the inverse of the FIM, the CRLB for estimating $\theta$ is derived in the following theorem.

\begin{thm}\label{thm:CRLB}
The expression for the CRLB of the target angle $\theta$ in the considered system is given as
\begin{align}
\label{crlb}
\mathrm{CRLB}(\theta) = \frac{\sigma_\mathrm{s}^2}{2L |\beta_\mathrm{s}|^2 \left( \| \dot{\mathbf{a}} \|^2 \, (\mathbf{b}^{\H} \mathbf{R}_\mathrm{x} \mathbf{b}) + N_\mathrm{t} \, (\dot{\mathbf{b}}^{\H} \mathbf{R}_\mathrm{x} \dot{\mathbf{b}}) \right)},
\end{align}
where 
\begin{align}
    \|\dot{\mathbf{a}}\|^2 &= \pi^2 \cos^2(\theta) \frac{N_\mathrm{t}(N_\mathrm{t}^2 - 1)}{12}\\\dot{\mathbf{a}}&= j \pi \cos\theta \cdot \mathbf{z}_\mathrm{t} \circ \mathbf{a}, \\
    \dot{\mathbf{b}}&= j \pi \cos\theta \cdot \mathbf{z}_\mathrm{r} \circ \mathbf{b},\\
    \mathbf{z}_\mathrm{x} &= \left[-\frac{N_\mathrm{x}-1}{2}, \dots, 0, \dots, \frac{N_\mathrm{x}-1}{2} \right]^\T,
\end{align}
for $\mathrm{x} \in \{\mathrm{t}, \mathrm{r}\}$. 
\begin{IEEEproof}
See Appendix~\ref{app:crlb}.
\end{IEEEproof}
\end{thm}

\begin{remark}
From \eqref{crlb}, it is evident that the denominator of the CRLB depends on the number of transmit antennas $N_\mathrm{t}$ and radar cross section $\beta_\mathrm{s}$. Thus, increasing $N_\mathrm{t}$ leads to a decrease in the CRLB and improves the estimation accuracy. In contrast, a reduction in the complex coefficient of the target, i.e., $\beta_\mathrm{s}$ results in an increased CRLB, indicating degraded estimation performance.  
\end{remark}

 \begin{figure*}[t!]
\small
    \centering
    \hspace{-3mm}
{\includegraphics[scale=0.41]{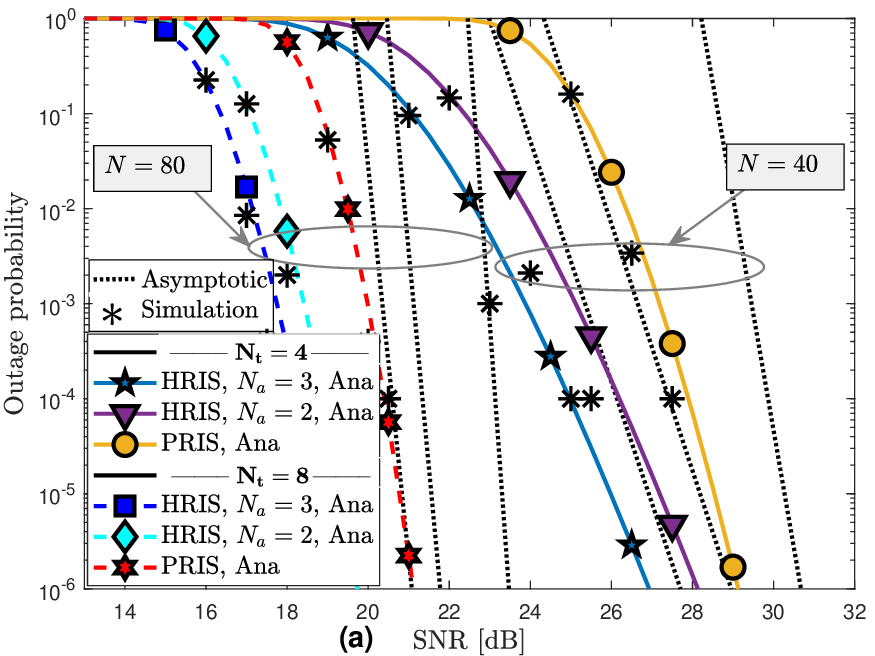}
\label{fig:outage_a}}
     \hspace{-4mm} 
    { \includegraphics[scale=0.41]{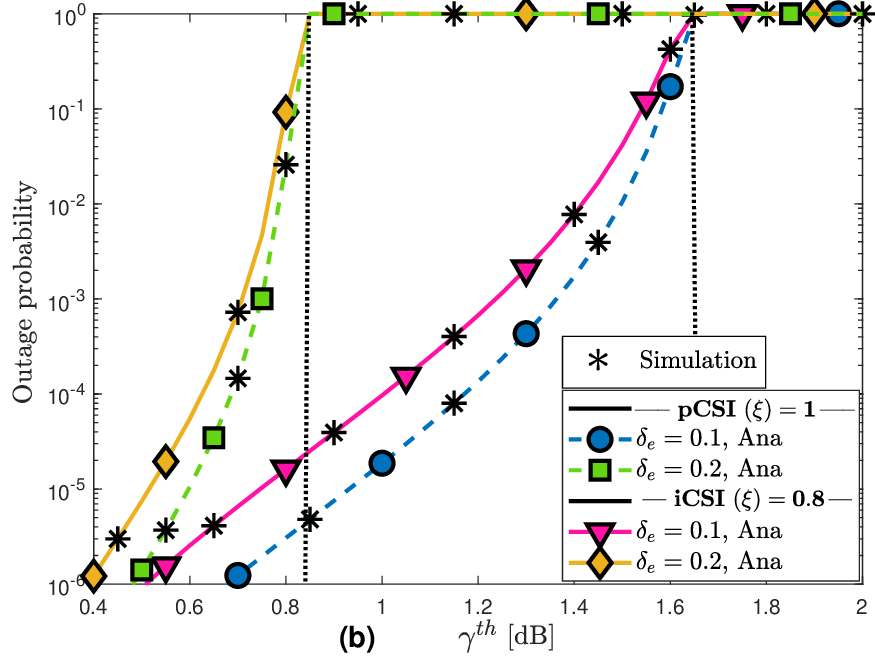}
    \label{fig:outage_b}
    }
       \hspace{-4mm} 
    { \includegraphics[scale=0.41]{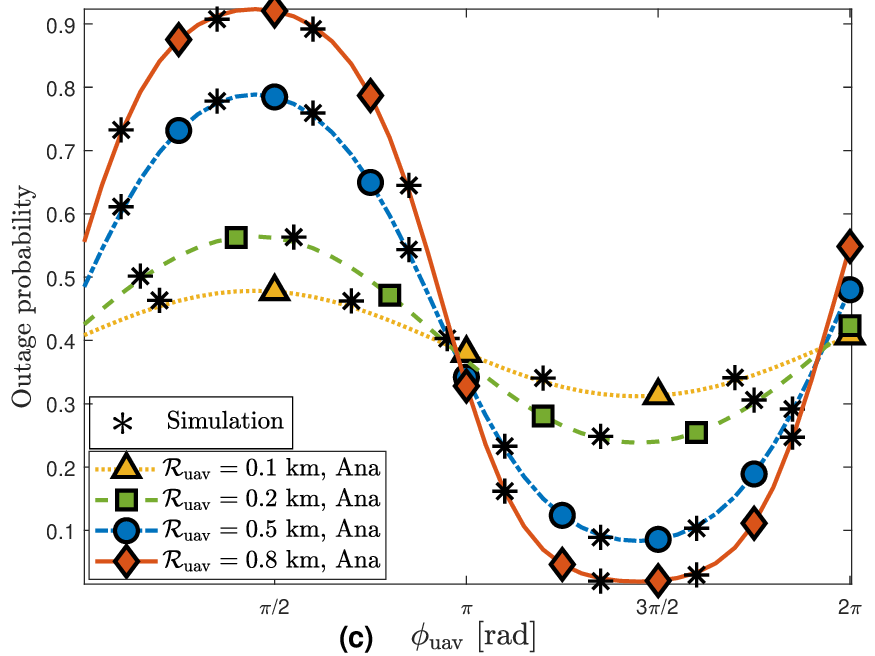}
    \label{fig:outage_c}
    }
\caption{\small{Impact of: (a) $N_\mathrm{t}$, $N$, and $N_\mathrm{a}$, (b) SINR threshold, and (c) {UAV parameters on the outage performance of the system.}}}
    \label{fig:outage} 
\end{figure*}

\begin{figure*}[t!]
\small
    \centering
    \hspace{-2mm}
{\includegraphics[scale=0.41]{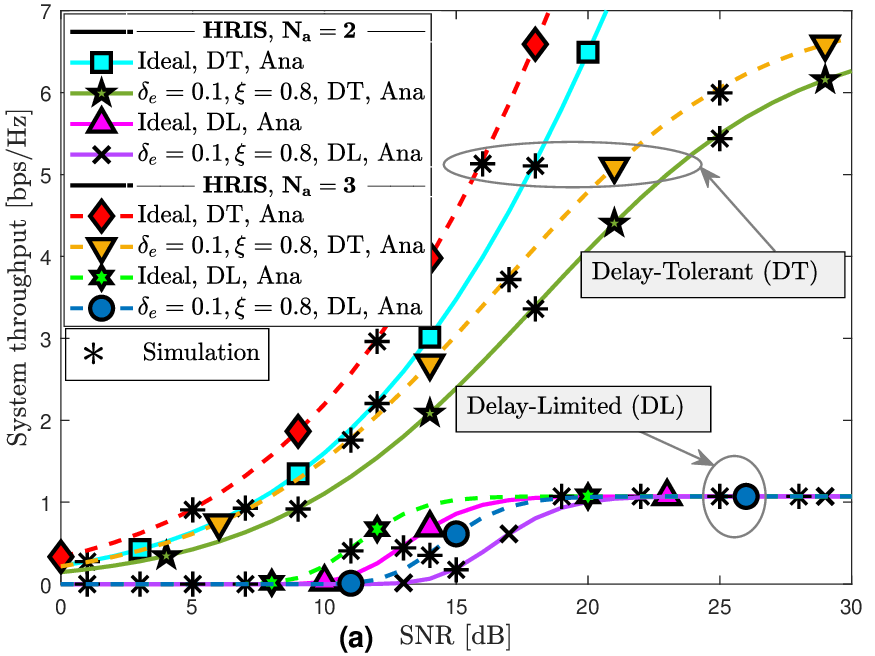}}
     \hspace{-3mm} 
    { \includegraphics[scale=0.41]{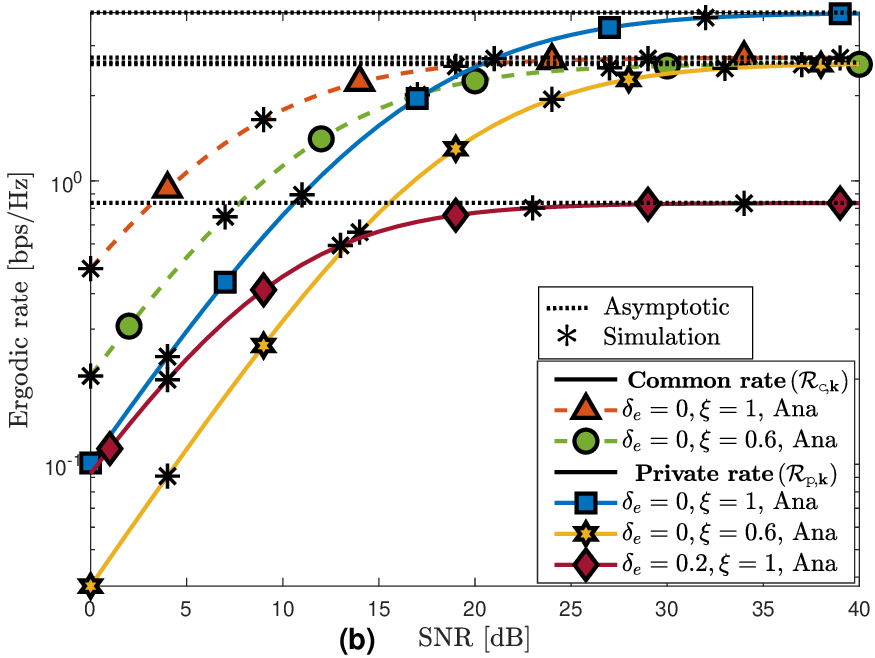}}
       \hspace{-3mm} 
    { \includegraphics[scale=0.41]{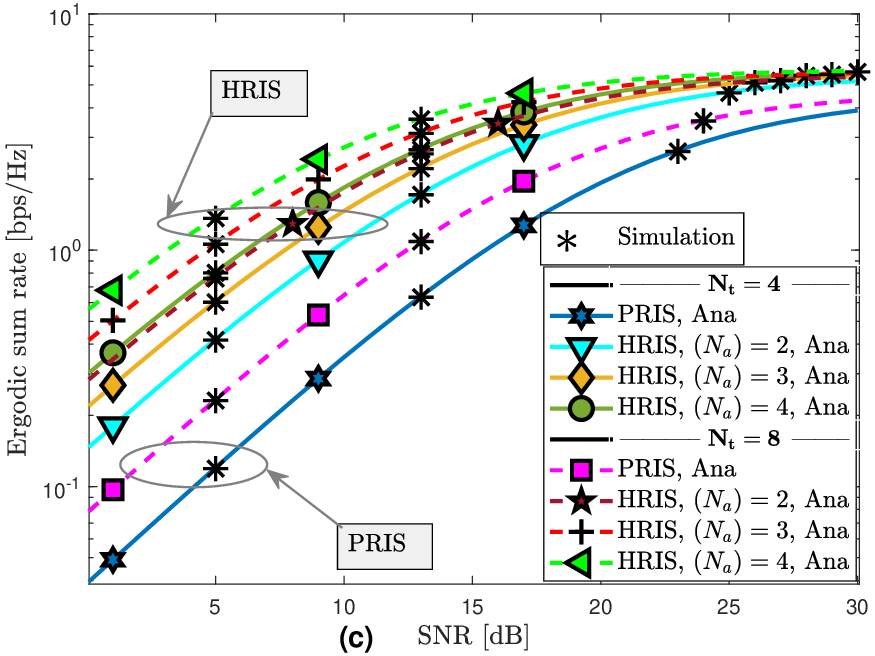}}
\caption{\small{Communication performance: (a) throughput under DT and DL modes, (b) ER with iSIC and iCSI, (c) ergodic sum rate versus SNR.}}
    \label{fig:communication_overall} 
\end{figure*}

\begin{table}[t]
\centering
\caption{{Simulation Parameters}}
\label{tab:sim_params}
\renewcommand{\arraystretch}{0.9}
\setlength{\tabcolsep}{3pt}
{\begin{tabular}{|l|c|}
\hline
\multicolumn{1}{|c|}{\textbf{Parameter}} & \multicolumn{1}{c|}{\textbf{Value}} \\
\hline\hline
Source position $(\mathcal{X}_s,\mathcal{Y}_s,\mathcal{Z}_s)$ & $(-2,-6,0)$~km \\
\hline
UAV parameters $(\mathcal{R}_\mathrm{uav},\mathcal{H}_\mathrm{uav},\phi_\mathrm{uav})$ & $(0.8~\mathrm{km},0.5~\mathrm{km},\pi/2)$ \\
\hline
User positions $(\mathcal{X}_{i_1},\mathcal{Y}_{i_1},\mathcal{Z}_{i_1})$ & $(2.5,-2,0)$~km \\
\hline
\phantom{User positions }$(\mathcal{X}_{i_2},\mathcal{Y}_{i_2},\mathcal{Z}_{i_2})$ & $(2,0,0)$~km \\
\hline
Path-loss coefficients $(a_\mathrm{x},b_\mathrm{x},c_\mathrm{x},g_\mathrm{x},\epsilon_\mathrm{x})$~\cite{Solanki_TVT} & $(-1.5,3.5,20,0.5,2.7)$ \\
\hline
Transmit/receive antennas $N_\mathrm{t}=N_\mathrm{r}$ & $4$ \\
\hline
Number of HRIS elements $N$ & $64$ \\
\hline
Common power coefficient $\alpha_\mathrm{c}$ & $0.6$ \\
\hline
SIC error level $\delta_\mathrm{e}$ & $0.2$ \\
\hline
Active-element gain $\eta_n,\ \forall n\in\mathcal{I}_\mathrm{a}$ & $3$~dB \\
\hline
$(m_\mathrm{sr},\Omega_\mathrm{sr})$, $(m_{\mathrm{R},k},\Omega_{\mathrm{R},k})$ & $(2,1)$, $(1,1)$ \\
\hline
Noise variances $\sigma_\mathrm{a}^2=\sigma_k^2$~\cite{nguyen2025performance} & $1$ \\
\hline
Target rates $(R^\mathrm{c},R^{\mathrm{p},k})$~\cite{li2024covert} & $(0.21,0.21)$~bps/Hz \\
\hline
Reference distance $d_0$~\cite{li2024covert} & $1$ \\
\hline
GCQ parameters $T_1=T_2$~\cite{li2024covert} & $20$ \\
\hline
\end{tabular}
}
\end{table}
\section{Numerical and Simulation Results}\label{num}
In this section, we validate the analytical results through extensive Monte Carlo simulations performed over $10^{4}$ channel realizations. Unless otherwise specified, the simulation parameters are summarized in Table~\ref{tab:sim_params}. {
The index set $\mathcal{I}_\mathrm{a}$ is randomly generated by selecting $N_\mathrm{a}$ elements from the $N$ HRIS elements \cite{nguyen2022spectral}. This configuration is then kept fixed throughout all numerical results.
}
We assume $\beta_\mathrm{s}=\frac{\lambda}{8d_\mathrm{s}^2}\sqrt{\frac{(G_\mathrm{t}G_\mathrm{r}\sigma)}{{\pi^3}}}$, where $\lambda=5.17$~cm, $G_\mathrm{t}=G_\mathrm{r}=2$~dB, $\sigma=0\ \text{dBsm} $ \cite{al_dweik}, and $d_\mathrm{s}$ is the distance between $S$ and $T$.

\subsection{Outage Probability}

Fig.~\ref{fig:outage}(\textcolor{red}{a}) shows OP vs. SNR for HRIS and PRIS under varying $N_\mathrm{t}$, $N$, and $N_\mathrm{a}$. {It can be observed that the analytical curves follow the same trend as the simulation results. For $N=40$, we note that the slight gap between the analytical and simulation results is due to the Gamma/moment-matching characterization of the cascaded channel. As expected, the analytical and simulation results align well when $N$ increases from $40$ to $80$, supporting the accuracy of Theorem~\ref{thm:out_accurate} in the large-HRIS regime.}
 At high SNRs, the asymptotic expressions coincide with the simulation results, validating \eqref{outage_asy_eqn}. {Increasing the number of HRIS elements from $N=40$ to $80$ consistently lowers the OP. This is reasonable because a larger HRIS aperture provides higher beamforming gain, which strengthens the cascaded channel and improves reception reliability.} Similarly, increasing $N_\mathrm{t}$ reduces OP by offering a higher array gain and spatial diversity, thereby mitigating deep fades. Finally, we observe that even with a few active elements ($N_\mathrm{a}$) the HRIS outperforms the PRIS with the same transmit power at the source. This is because active elements in the HRIS amplify and preprocess the incident signal before reflection which improves the channel gain and reliability at a given SNR.

Fig.~\ref{fig:outage}(\textcolor{red}{b}) shows OP vs. the SINR threshold $\gamma^{\mathrm{th}}$ for different CSI ($\xi$) and iSIC  ($\delta_\mathrm{e}$) levels. We observe that changing $\xi$ from pCSI ($\xi=1$) to iCSI ($\xi=0.8$) does not change the saturation threshold, i.e., OP ceiling point. In contrast, increasing the SIC error from $\delta_\mathrm{e}=0.1$ to $\delta_\mathrm{e}=0.2$ shifts the saturation threshold from $\gamma^{\mathrm{th}}=1.67$ to $\gamma^{\mathrm{th}}=0.85$, demonstrating the strong effect of residual interference. For a fixed $\delta_\mathrm{e}=0.1$, the loss due to CSI inaccuracy is more visible at small $\gamma^{\mathrm{th}}$. However, for $\delta_\mathrm{e}=0.2$, this difference decreases because the system is interference limited, thus, the extra impact of iCSI is smaller. These observations are in line with the discussion in Remark~\ref{threshold_cond}.

Fig.~\ref{fig:outage}(\textcolor{red}{c}) shows the OP versus the UAV's elevation angle $\phi_\mathrm{uav}$ for different values of radius $\mathcal{R}_\mathrm{uav}$. It can be seen that the OP peaks near $\phi_{\mathrm{uav}} \approx \pi/2$, i.e., when the UAV is farthest from the user and the path-loss is largest. In contrast, the OP is lowest at $\phi_{\mathrm{uav}} \approx 3\pi/2$, i.e., when moving toward the user and benefits from a shorter link distance. For $\phi_{\mathrm{uav}} = \pi/2$, increasing $R_{\mathrm{uav}}$ worsens the OP, while for $\phi_{\mathrm{uav}} = 3\pi/2$, a larger $R_{\mathrm{uav}}$ reduces the OP by allowing better relay positioning and stronger received signals.

\begin{figure*}[t!]
\small
    \centering
    \hspace{-2mm}
{\includegraphics[scale=0.41]{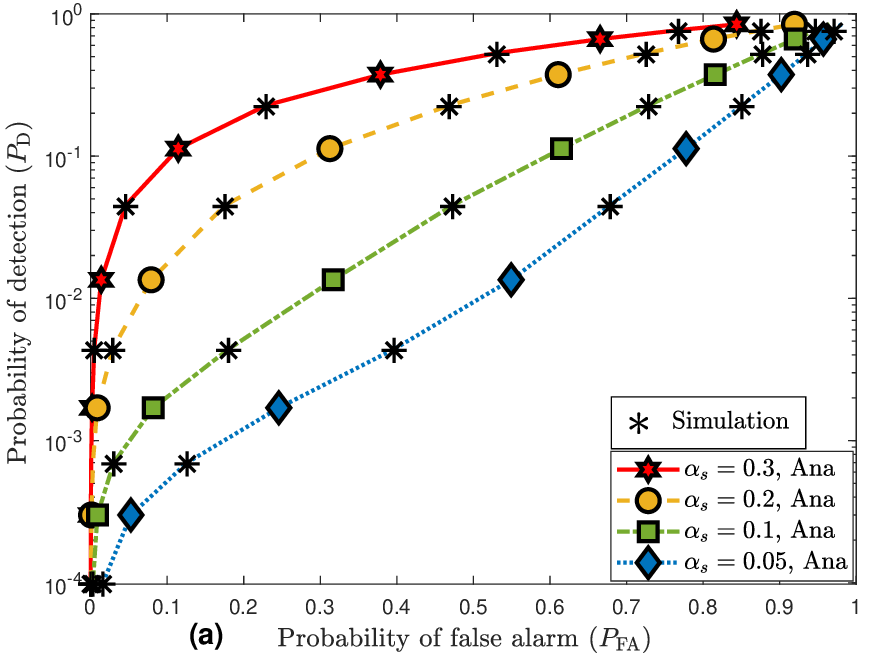}}
     \hspace{-3mm} 
    { \includegraphics[scale=0.41]{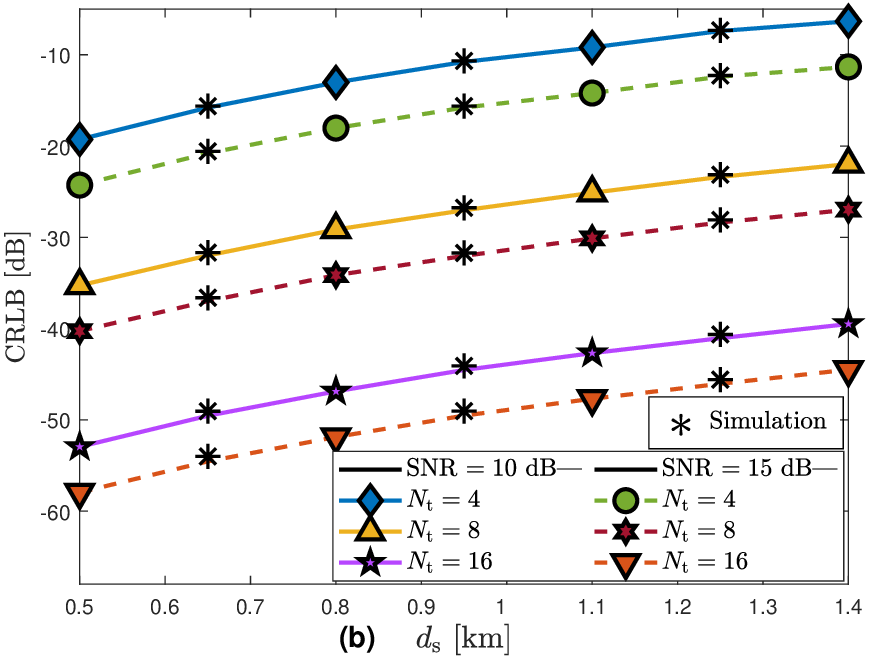}}
       \hspace{-3mm} 
    { \includegraphics[scale=0.41]{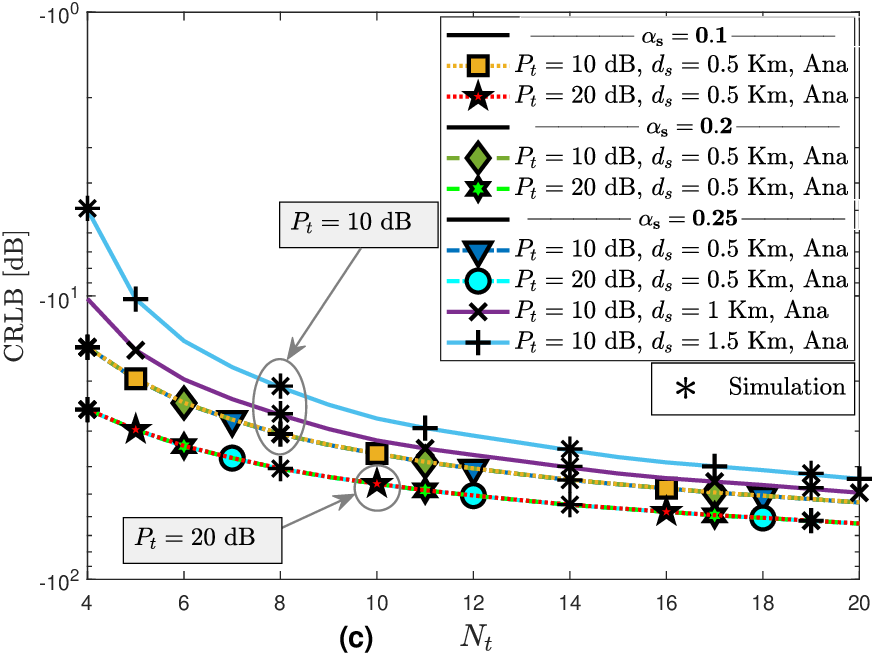}}
\caption{\small{Sensing performance: (a) $P_\mathrm{D}$ vs. $P_\mathrm{FA}$ for varying $\alpha_\mathrm{s}$, (b) CRLB vs. target distance with different $N_\mathrm{t}$, (c) CRLB vs. $N_\mathrm{t}$ for different values of $P_\mathrm{t}$ and target distance.}} 
    \label{fig:sensing} 
\end{figure*}

\subsection{Communication Sum-Rate}

Fig.~\ref{fig:communication_overall}(\textcolor{red}{a}) depicts the impact of iCSI and iSIC on the system throughput for various values of the SNR. It is seen that the throughput improves with a higher SNR owing to the increased transmit power. In DL mode, the throughput initially increases and eventually saturates. This is because strict latency constraints can limit the achievable rate, irrespective of the signal strength. In contrast, the DT mode consistently achieves a higher throughput by averaging the channel fluctuations over longer durations, effectively utilizing the available SNR. It is also observed that increasing $N_\mathrm{a}=2$ to $N_\mathrm{a}=3$ enhances the throughput in both modes by providing improved amplification. Furthermore, under non-ideal conditions, i.e., $\delta_\mathrm{e}=0.1$, $\xi=0.8$ with $N_\mathrm{a}=2$, the throughput drops by approximately $31\%$ in DT mode and nearly $80\%$ in DL mode compared to ideal scenarios. This reveals that the DL mode is more sensitive to CSI errors and SIC imperfections, while DT is more resilient due to its flexibility in handling the channel variations.

Fig.~\ref{fig:communication_overall}(\textcolor{red}{b}) shows the ER versus SNR for both common and private messages under varying CSI inaccuracy ($\xi$) and SIC error ($\delta_\mathrm{e}$). We first observe that the analytical results align well with the simulation results, validating Theorem~\ref{thm:rate_accurate}. 
It can be seen that for the common rate, with $\delta_\mathrm{e} = 0$, the performance gap between $\xi = 1$ and $0.6$ is significant at low SNR. This suggests that the decoding of the common message becomes more sensitive to
CSI degradation in noise-limited regime. However, it converges to the ideal CSI case in the high-SNR regime. For the private rate, the decrease in ER is greater at high SNRs than at low SNRs. For example, CSI degradation causes the private rate to degrade by $35\%$ at $27$~dB SNR and $62\%$ at $5$~dB SNR. Furthermore, it is evident that the impact of 
the SIC error i.e., $\delta_\mathrm{e} = 0.2$ is more prominent at high SNRs. This is because, at high SNRs, AWGN noise becomes negligible, and residual interference due to the iSIC dominates the system performance. The inability to fully cancel this interference limits ER. In contrast, at low SNRs, noise masks this effect, making the impact of the SIC error less pronounced. In addition, the ER saturates at high SNRs, as discussed in Remark~\ref{ER_asy_eqn}.

Fig.~\ref{fig:communication_overall}(\textcolor{red}{c}) illustrates the ergodic sum rate versus SNR for the PRIS and HRIS systems under varying numbers of transmit antennas (${N_\mathrm{t}}$) and ARIS elements ($N_\mathrm{a}$). As can be seen, with only a few active elements, the HRIS can consistently outperform PRIS for the entire considered range of SNRs. For example with just $2$ active elements, the  ER of HRIS is $89\%$ higher than that of PRIS for ${N_\mathrm{t}}=4$ and SNR $=20$~dB. Furthermore, it is evident that by increasing ${N_\mathrm{t}}$ the ergodic sum rate significantly increases, due to the same reason as highlighted in Fig.~\ref{fig:outage}(\textcolor{red}{a}). These trends confirm that HRIS with a few active elements and higher ${N_\mathrm{t}}$ provides significant performance gains over PRIS at very-low SNRs. However, at high SNRs, the ergodic sum rate saturates because of the interference caused by the sensing target and iSIC.

\subsection{Sensing Performance Metrics}
Fig.~\ref{fig:sensing}(\textcolor{red}{a}) shows the variation of the PoD, i.e., $P_{\mathrm{D}}$ with the PoFA, i.e., $P_{\mathrm{FA}}$ for different values of the sensing power coefficient $\alpha_\mathrm{s}$. It is seen that for a given $P_{\mathrm{FA}}$, increasing $\alpha_\mathrm{s}$ significantly improves $P_{\mathrm{D}}$ because more transmit power is allocated to sensing, enhancing the echo signal strength and target detectability. Furthermore, at $P_{\mathrm{FA}} = 0.2$, increasing $\alpha_\mathrm{s}$ from $0.05$ to $0.3$ yields several orders of magnitude improvement in $P_{\mathrm{D}}$. Conversely, smaller $\alpha_\mathrm{s}$ values limit the sensing power, resulting in poorer detection, particularly in the low $P_{\mathrm{FA}}$ regime.

Fig.~\ref{fig:sensing}(\textcolor{red}{b}) shows that the CRLB increases with the target distance $d_\mathrm{s}$ due to the higher path loss, thereby reducing the sensing accuracy. It can be seen that a higher SNR improves the CRLB performance. In addition, increasing the number of transmit/receive antennas ($N_\mathrm{t}=N_\mathrm{r}$) further reduces the CRLB by providing greater beamforming gain, which concentrates energy toward the target and enhances the received echo strength. For example, at $d_\mathrm{s} = 0.8$ km and SNR $=10$ dB, increasing $N_\mathrm{t}$ from $4$ to $8$ improves the CRLB performance by nearly $97.5\%$.

Fig.~\ref{fig:sensing}(\textcolor{red}{c}) shows CRLB versus the number of transmit/receive antennas $N_\mathrm{t}=N_\mathrm{r}$ for various values of  transmit power $P_\mathrm{t}$, sensing power factors $\alpha_\mathrm{s}$, and target distances $d_\mathrm{s}$. As seen, the CRLB decreases monotonically with $N_\mathrm{t}$, highlighting the improved sensing performance due to the increased array gain. It is also observed that a higher $P_\mathrm{t}$, i.e., $10$ dB to $20$ dB further lowers the CRLB by increasing the radar SINR. In contrast, a larger $d$ increases the CRLB, since the path-loss attenuates the received echo signal. Furthermore, increasing $\alpha_\mathrm{s}$ improves the CRLB performance by allocating more power to the sensing sub-system.

\subsection
{Sensing and Communication Tradeoff}
\begin{figure*}[t!]
\small
    \centering
    \hspace{-2mm}
{\includegraphics[scale=0.41]{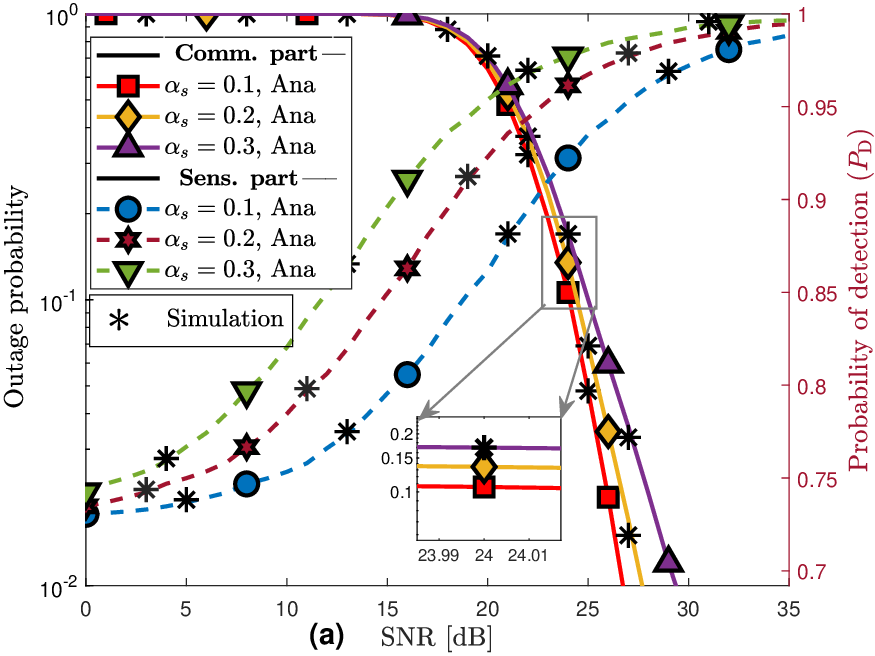}}
     \hspace{-3mm} 
    { \includegraphics[scale=0.41]{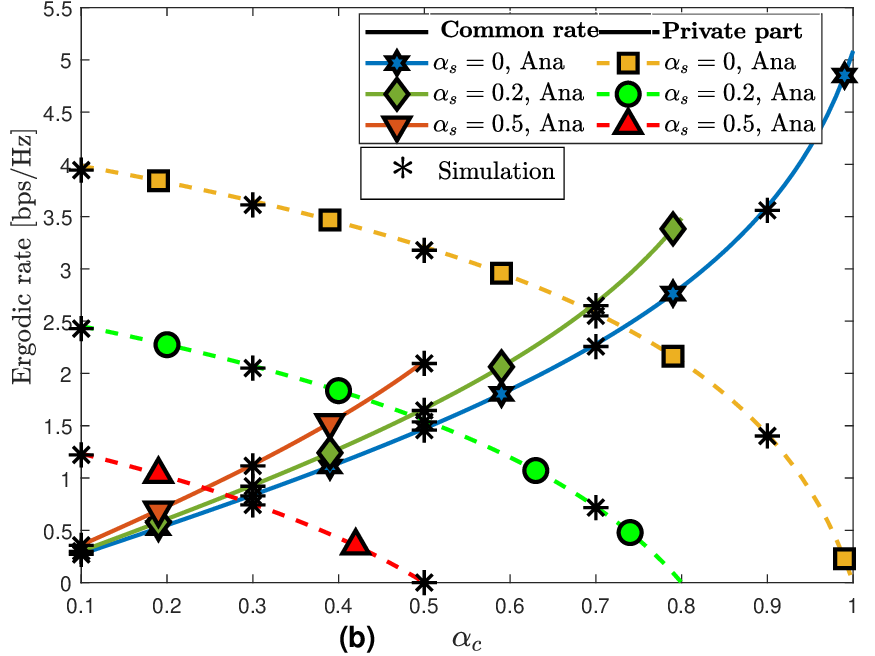}}
       \hspace{-3mm} 
    { \includegraphics[scale=0.41]{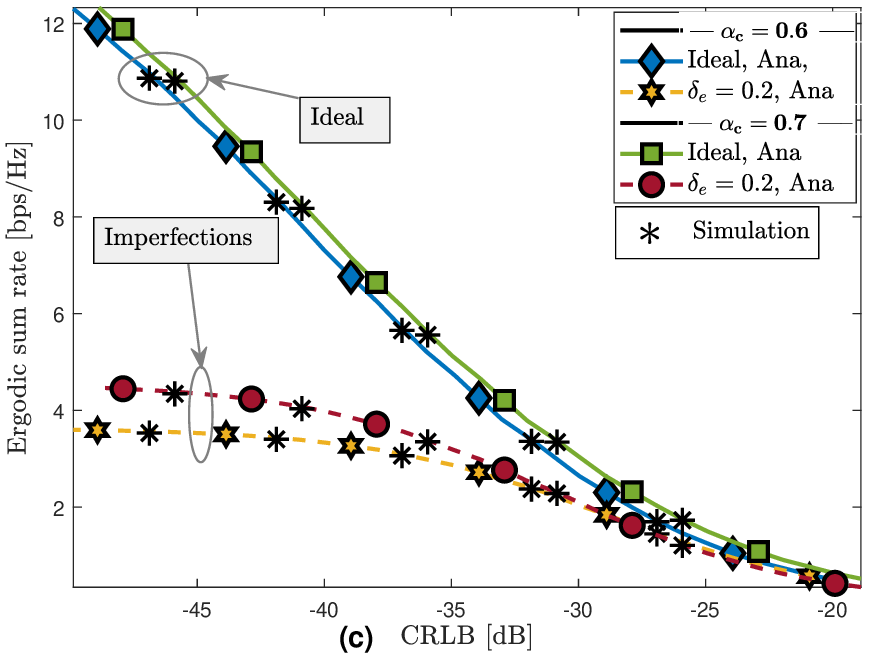}}
\caption{\small{Tradeoffs in sensing-communication performance: (a) OP vs. PoD, (b) ER vs. $\alpha_c$ with varying $\alpha_\mathrm{s}$, (c) ergodic sum rate vs. CRLB.}}
    \label{fig:tradeoff} 
\end{figure*}

{Fig.~\ref{fig:tradeoff} shows the relationships among the considered communication and sensing metrics. In particular, the OP--PoD curve shows the tradeoff between communication reliability and sensing detection reliability, while the ergodic sum-rate--CRLB curve shows the tradeoff between communication spectral efficiency and sensing estimation accuracy. These metrics are coupled through the power allocation constraint $\alpha_{\mathrm{c}}+\sum_{k=1}^{K}\alpha_k+\alpha_{\mathrm{s}}=1$.} More specifically,
in Fig.~\ref{fig:tradeoff}(\textcolor{red}{a}) we illustrate the tradeoff between OP and PoD in the considered system for various values of SNRs and  $\alpha_\mathrm{s}$. As discussed, $\alpha_\mathrm{s}$ is constrained as $\alpha_{\mathrm{c}} + \sum_{k=1}^{K} \alpha_k + \alpha_{\mathrm{s}} = 1$. This  implies that allocating more power to sensing reduces the available power for communication signals. Thus, the OP increases with higher $\alpha_\mathrm{s}$, and less transmit power is allocated to common and private messages, leading to degraded signal quality at the users. In contrast, the PoD improves with increasing $\alpha_\mathrm{s}$ as allocating more power for sensing enhances target illumination and improves detection accuracy. This trend highlights the fundamental power-sharing tradeoff in ISAC, where improved sensing reliability comes at the cost of reduced communication performance and vice versa.

Fig.~\ref{fig:tradeoff}(\textcolor{red}{b}) shows that increasing $\alpha_\mathrm{c}$ improves the common rate by allocating more power to the common stream, thereby enhancing the decoding reliability of the communication user. In contrast, the private rate decreases because private power coefficient decreases with the increase in $\alpha_\mathrm{c}$. Furthermore,  a higher value of $\alpha_\mathrm{s}$ reduces the common and private ERs, by allocating less power for communication. This reduction is more pronounced for the private ER, as its smaller power-allocation coefficient makes it more sensitive to reductions in the allocated power. Therefore, selecting a balance between $\alpha_\mathrm{c}$ and $\alpha_\mathrm{s}$ is crucial to maintain both high communication performance and sensing capability.

In Fig.~\ref{fig:tradeoff}(\textcolor{red}{c}), we plot the ergodic sum rate versus CRLB, both evaluated over the same SNR range, i.e., $0- 30$~dB. We observe that as $\alpha_\mathrm{c}$ increases from $0.6$ to $0.7$, with private power coefficient $\alpha_k$ kept constant, the sum rate increases. On the other hand, the CRLB shifts to the right, indicating reduced sensing accuracy. This is reasonable  because increasing $\alpha_\mathrm{c}$ reduces the  power allocated to sensing subsystem, thereby degrading the CRLB performance. Furthermore, the presence of iSIC errors, i.e.,  $\delta_\mathrm{e} = 0.2$ significantly degrades the ergodic sum rate compared to the ideal case, owing to a similar reason as discussed in Fig.~\ref{fig:outage}(\textcolor{red}{b}). {Furthermore, we observe that the curves corresponding to $\alpha_{\mathrm{c}}=0.6$ and $\alpha_{\mathrm{c}}=0.7$  intersect under iSIC. This is due to the common/private power-splitting tradeoff in RSMA with iSIC. More specifically, from Fig.~\ref{fig:tradeoff}(\textcolor{red}{b}), we observe that increasing $\alpha_{\mathrm{c}}$ improves the common-stream rate but reduces the private-stream rate. This is because a larger $\alpha_{\mathrm{c}}$ leaves less power for the private streams and also increases the residual common-stream interference term $\delta_{\mathrm{e}}\alpha_{\mathrm{c}}\beta_k\rho_{\mathrm{t}}U_k$ in the denominator of the private-stream SINR in~\eqref{gamma_pk}. 
Therefore, the ergodic sum rate can exhibit a non-monotonic behavior with respect to $\alpha_{\mathrm{c}}$ under iSIC.
}

 \begin{figure}[t!]
	\centering
\includegraphics[height=7cm,width=8.5cm]{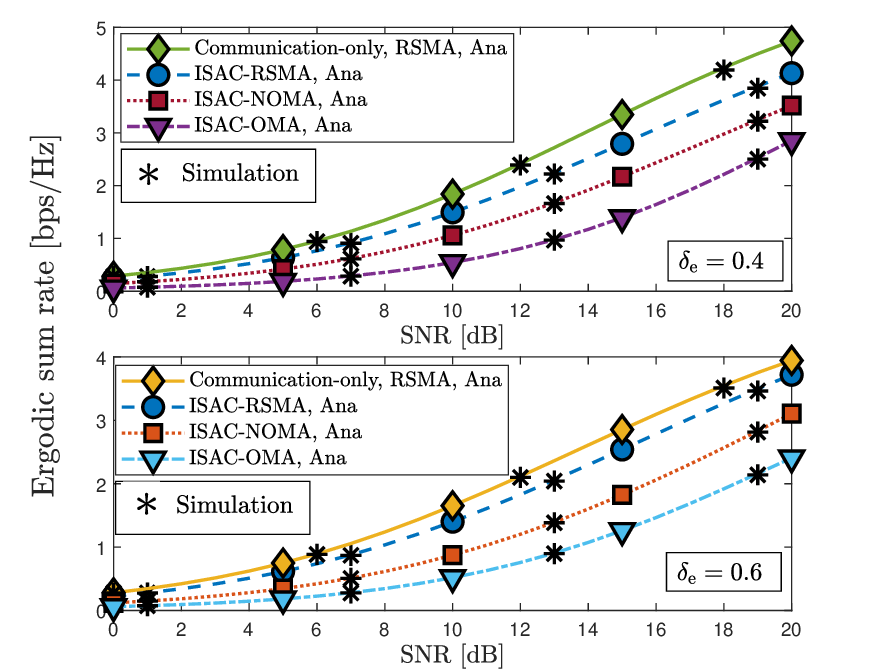}	\caption{\small Performance comparison with benchmark multiple access schemes.}
	\label{compare}
\end{figure}
\subsection
{Performance Comparison with Benchmark Schemes}
In Fig. \ref{compare}, we compare the ergodic sum rate performance of the proposed RSMA-based ISAC system with the NOMA- and OMA-based ISAC systems. {
The OMA baseline is implemented as a special case of the considered RSMA protocol by scheduling only one user at a time~\cite{clerckx2023primer}. Specifically, when user $k$ is served, the common-stream power coefficient and the power coefficients of the private streams intended for the other users are set to zero, i.e., $\alpha_{\mathrm{c}}=0$ and $\alpha_j=0$, $\forall j\neq k$. The scheduled user is allocated the remaining communication power, i.e., $\alpha_k=1-\alpha_{\mathrm{s}}$. For a fair comparison, the same $\alpha_{\mathrm{s}}$ is used as that for the RSMA and NOMA schemes.
}
Note that in Fig. \ref{compare}, the proposed UAV-mounted HRIS-assisted ISAC system employing RSMA is abbreviated as “ISAC-RSMA” for brevity. It is observed that all three multiple access scheme-based ISAC systems exhibit comparable performance at low SNRs, whereas the proposed RSMA-based ISAC system achieves the highest ergodic sum rate, with a noticeable gain beyond an SNR of $10$ dB. We further include the communication-only RSMA HRIS-UAV benchmark as an upper bound. It can be seen that the proposed RSMA-based ISAC system remains the closest to the communication-only benchmark. This indicates that RSMA-based ISAC system retains a larger fraction of the communication performance while satisfying the sensing requirements. Moreover, as the SIC error increases, the gap between the communication-only RSMA and RSMA-based ISAC curves becomes smaller. This highlights the robustness of the proposed RSMA-based ISAC system to SIC imperfections. In contrast, the NOMA- and OMA-based equivalent ISAC systems exhibit nearly unchanged gaps relative to the communication-only benchmark as the SIC error increases.

 \begin{figure}[t!]
	\centering
\includegraphics[height=6cm,width=8.5cm]{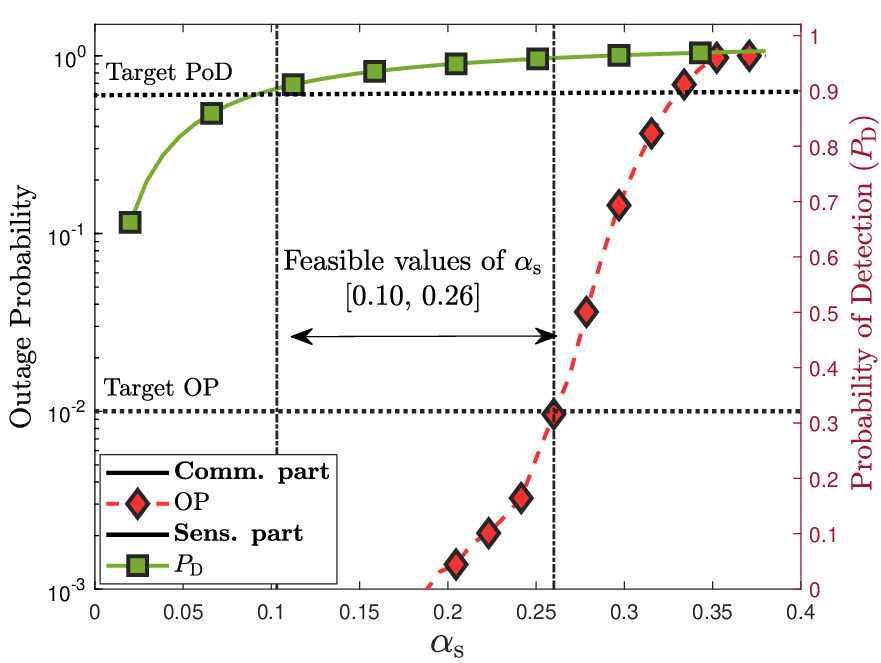}	\caption{\small {Feasible values of $\alpha_\mathrm{s}$ for a given target OP and PoD.}}
	\label{alpha_s}
\end{figure}
\subsection{Feasible $\alpha_\mathrm{s}$ Values for Balancing OP and PoD }
{Fig.~\ref{alpha_s} illustrates the feasible range of the sensing power allocation coefficient $\alpha_\mathrm{s}$ by jointly considering the OP and PoD. As observed, for the considered requirements $\mathcal{P}_{\mathrm{o}_k}\leq 10^{-2}$ and $P_{\mathrm{D}}\geq 0.9$, the feasible values of $\alpha_\mathrm{s}$ lie in the range $0.103\leq\alpha_\mathrm{s}\leq0.251$. The lower bound is determined by the PoD requirement, since a smaller $\alpha_\mathrm{s}$ allocates insufficient power to the sensing stream and yields $P_{\mathrm{D}}<0.9$. In contrast, the upper bound is determined by the OP requirement, since a larger $\alpha_\mathrm{s}$ reduces the power allocated to the common and private messages, thereby increasing $\mathcal{P}_{\mathrm{o}_k}$. Hence, selecting $\alpha_\mathrm{s}$ within this range ensures that both the OP and PoD requirements are satisfied. Similarly, for other required OP and PoD values, the corresponding feasible range of $\alpha_\mathrm{s}$ can be obtained using the derived expressions for the communications and sensing performance metrics.}

\begin{figure}[t!]
	\centering
\includegraphics[height=6cm,width=8.5cm]{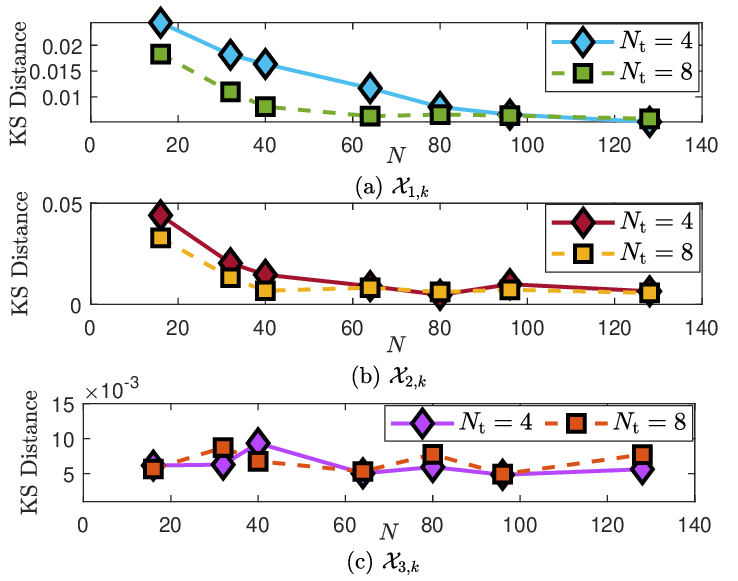}	\caption{\small {KS-distance versus the number of HRIS elements $N$ for the RVs $\mathcal{X}_{1,k}$, $\mathcal{X}_{2,k}$, and $\mathcal{X}_{3,k}$.}}
	\label{fig:ks_validation}
\end{figure}
\subsection{Validation of Gamma-Based Approximations}
{To validate the Gamma approximations presented in Lemma~\ref{lemma:RV}, we show the Kolmogorov--Smirnov (KS) distance, i.e., $D_{\mathrm{KS},\mathcal{X}_{i,k}}\triangleq\sup_{x\geq0}\left|F_{\mathcal{X}_{i,k}}^{\mathrm{emp}}(x)-F_{\mathcal{X}_{i,k}}^{\mathrm{ana}}(x)\right|$, for $i\in\{1,2,3\}$. Here, $F_{\mathcal{X}_{i,k}}^{\mathrm{emp}}(x)$ is obtained from Monte Carlo realizations, whereas $F_{\mathcal{X}_{i,k}}^{\mathrm{ana}}(x)$ is obtained from the Gamma shape and rate parameters given in Lemma~\ref{lemma:RV}. A smaller KS distance indicates a closer agreement between the empirical and analytical distributions. As shown in Fig.~\ref{fig:ks_validation}, the KS distances decreases with increase in $N$. In particular, at $N=64$ and $N_{\mathrm{t}}=4$, the KS distances of $\mathcal{X}_{1,k}$, $\mathcal{X}_{2,k}$, and $\mathcal{X}_{3,k}$ are approximately $1.2\times10^{-2}$, $8\times10^{-3}$, and $5\times10^{-3}$, respectively, confirming the accuracy of the adopted approximations. Moreover, the KS distances of $\mathcal{X}_{3,k}$ remain nearly unchanged as $N$ increases. This is because $\mathcal{X}_{3,k}$ depends predominantly on the active-HRIS elements with $N_{\mathrm{a}}\ll N$.}

\section{Conclusion}\label{conclu}
We analyzed the performance of a UAV-mounted HRIS-assisted monostatic ISAC system employing RSMA transmission scheme in the presence of practical impairments, i.e., iCSI and iSIC. We derived accurate and asymptotic expressions for the OP and ER under Nakagami-$m$ fading channels.
In addition, we provided the expressions for the system throughput for the DT and DL transmission modes. To assess the performance of the sensing subsystem, we provided the expressions for the $P_\mathrm{D}$, $P_\mathrm{FA}$, and CRLB.
We showed that the HRIS-assisted system can outperform its PRIS counterpart by $97\%$ in terms of OP and by $89\%$ in terms of ER, with only $2$ active elements. Moreover, we provided several insights, including the diversity order and coding gain, the impact of iSIC and CSI on OP, ER, and throughput. Our analysis revealed that both the DL and DT modes benefit from the HRIS, whereas the DT mode suffers significantly from CSI and SIC imperfections. 
We also demonstrated the critical impact of the power allocation coefficients on the tradeoff between communication and sensing performance. Finally, we corroborated the analytical findings through extensive Monte Carlo simulations, which confirmed the performance gains of HRIS-UAV and RSMA in the proposed ISAC system. {Furthermore, deriving similar sensing metrics in multi-target ISAC scenarios, while accounting for clutter and mutual target interference, remains an important direction for future work.}


 \appendices
\section{Proof of Lemma~\ref{lemma:RV}}\label{app:lemma_rv}
\renewcommand{\theequation}{A.\arabic{equation}}
\setcounter{equation}{0}

We utilize the Gamma distribution to model the RVs $\mathcal{X}_{1,k} \triangleq \rho_\mathrm{t} \beta_k U_k,$ $\mathcal{X}_{2,k}  \triangleq \beta_k \alpha_\mathrm{s} \rho_\mathrm{t}Z_k,$ and
$\mathcal{X}_{3,k} \triangleq \tilde{\beta}_k V_k$. 
\begin{itemize}
    \item \textit{Compute $\mathcal{X}_{1,k}$:} We first compute $\mathbb{E}[U_k]$ as  $\mathbb{E}[U_k]=\mathbb{E}[\|\hat{\mathbf{h}}_{k}\|^2]$.
The scalar representation of $\|\hat{\mathbf{h}}_{k}\|^2$ is given as 
 $\|{\mathbf{h}}_{k}\|^2= \sum_{i=1}^{{N_\mathrm{t}}} |{{h}}_{k,i}|^2 $ with $
 |{{h}}_{k,i}|^2 = \Bigl\lvert
     \sum_{n=1}^{N} |\eta_n| h_{\mathrm{sr}_n,i} h_{\mathrm{r}k,n}
       e^{j\phi_{n}}
   \Bigr\rvert^{2}
$. We use Gamma distribution as $\lvert h_{k,i}\rvert^{2}
   \sim
   \mathrm{Gamma}\bigl(
      \alpha_{|h_{k,i}|^{2}},
      \beta_{|h_{k,i}|^{2}}
   \bigr)$ which requires the moments of $\|{\mathbf{h}}_{k}\|^2$ to be evaluated.
Based on the iCSI model in \eqref{icsi}, we have
\begin{align}
\label{hk_hat}
& \mathbb{E}[\|{\mathbf{\hat{h}}}_{k}\|^2]=\xi^{2} \mathbb{E}[\|{\mathbf{h}}_{k}\|^2] + (1-\xi^{2}) \mathbb{E}[\|{\mathbf{\bar{h}}}_{k}\|^2],
& \\
& \mathbb{E}[\|{\mathbf{\hat{h}}}_{k}\|^4]=\xi^{4} \mathbb{E}[\|{\mathbf{h}}_{k}\|^4] + (1-\xi^{2})^2 \mathbb{E}[\|{\mathbf{\bar{h}}}_{k}\|^4] \nonumber \\ 
& \hspace{16mm}+2 \xi^{2} (1-\xi^{2}) \mathbb{E}[\|{\mathbf{h}}_{k}\|^2] \mathbb{E}[\|{\mathbf{\bar{h}}}_{k}\|^2].
\label{hk_hat1}
\end{align}
From the statistics of a Gamma distributed RV we get
$\mathbb{E}\!\bigl[\lVert\mathbf h_k\rVert^{2}\bigr]
    ={N_\mathrm{t}}\,
        \frac{\alpha_{|h_{k,i}|^{2}}}{\beta_{|h_{k,i}|^{2}}}$ and $
\mathbb{E}\!\bigl[\lVert\mathbf h_k\rVert^{4}\bigr]={N_\mathrm{t}}\,
        \frac{\alpha_{|h_{k,i}|^{2}}
              (\alpha_{|h_{k,i}|^{2}}{N_\mathrm{t}}+1)}
             {\beta_{|h_{k,i}|^{2}}^{2}}$.
Besides, $\mathbb{E}\!\bigl[\lVert\bar{\mathbf h}_k\rVert^{2}\bigr]
      \overset{(a)}{=}
{N_\mathrm{t}}\,d_k^{-\tau}$ and $\mathbb{E}\!\bigl[\lVert\bar{\mathbf h}_k\rVert^{4}\bigr]
      \overset{(a)}{=}
{N_\mathrm{t}}({N_\mathrm{t}}+1)d_k^{-2\tau}$, where the equality $(a)$ follows due to the distribution of $\bar{\mathbf h}_k$ given in \eqref{icsi}. By using these moments of $\bar{\mathbf{h}}_k$ into 
\eqref{hk_hat}-\eqref{hk_hat1}, we get \eqref{hk_bold}-\eqref{hk_bold1}.
\item \textit{Compute $\mathcal{X}_{2,k}$:}
 We define $Z_k\triangleq|\hat{\mathbf{h}}_k^\H\mathbf{w}_\mathrm{s}|^2$, based on the iCSI model we can write $\hat{\mathbf{h}}_{k}^\H\mathbf{w}_\mathrm{s}
     =\xi\,\mathbf{h}_{k}^\H\mathbf{w}_\mathrm{s}
       +\sqrt{1-\xi^{2}}\;
        \bar{\mathbf{h}}_{k}^\H\mathbf{w}_\mathrm{s}
\triangleq
       \xi\,x
       +\sqrt{1-\xi^{2}}\;y$,
where $x=\mathbf{h}_{k}^\H\mathbf{w}_\mathrm{s}$ and $y=\bar{\mathbf{h}}_{k}^\H\mathbf{w}_\mathrm{s}$. Moreover,
for large $N$ and ${N_\mathrm{t}}$ we have $\mathbf{h}_k^\H \sim \mathcal{CN}(\mathbf{m}, \mathbf{R}_{\mathbf{h}_k})$, where $\mathbf{m}$ is the mean vector and $\mathbf{R}_{\mathbf{h}_k}$ is the covariance matrix of $\mathbf{h}_k$. 
Now, the statistics for the terms $x$ and $y$ can be given as,
\begin{align}
\mathbb{E}\bigl[|x|^2\bigr]
     &= \mathbb{E}\bigl[|\mathbf{h}_k^\H \mathbf{w}_\mathrm{s}|^2\bigr] \nonumber \\
     &= \mathbf{w}_\mathrm{s}^\H \mathbf{R}_{\mathbf{h}_k} \mathbf{w}_\mathrm{s}
        + |\mathbf{m}^\H \mathbf{w}_\mathrm{s}|^2, \\ \mathbb{E}\bigl[|y|^2\bigr]
     &= \mathbb{E}\bigl[|\bar{\mathbf{h}}_k^\H \mathbf{w}_\mathrm{s}|^2\bigr] = d_k^{-\tau} \lVert \mathbf{w}_\mathrm{s} \rVert^2.
     \end{align}
Using the above expressions for $\mathbb{E}\bigl[|x|^2\bigr]$ and $\mathbb{E}\bigl[|y|^2\bigr]$ we get,
\begin{align} \mathbb{E}\bigl[|\hat{\mathbf{h}}_k^\H \mathbf{w}_\mathrm{s}|^2\bigr]
  & = \xi^2 \left(
       \mathbf{w}_\mathrm{s}^\H \mathbf{R}_{\mathbf{h}_k} \mathbf{w}_\mathrm{s}
       + |\mathbf{m}^\H \mathbf{w}_\mathrm{s}|^2
     \right) \nonumber \\
     &
   + (1 - \xi^2)\, d_k^{-\tau} \lVert \mathbf{w}_\mathrm{s} \rVert^2. 
\end{align}
We next compute $\mathbb{E}\bigl[|\mathbf{h}_k^\H \mathbf{w}_\mathrm{s}|^4\bigr]$.
First, let us define $Z 
  = \xi\,x + \sqrt{1 - \xi^2} y$ and consider that $x \sim \mathcal{CN}(\mu_x, \sigma_x^2)$, where $\mu_x = \mathbf{m}^\H \mathbf{w}_\mathrm{s}$, $\sigma_x^2 = \mathbf{w}_\mathrm{s}^\H \mathbf{R}_{\mathbf{h}_k} \mathbf{w}_\mathrm{s}$, and $y \sim \mathcal{CN}(0, \sigma_y^2)$, with $\sigma_y^2 = d_k^{-\tau} \lVert \mathbf{w}_\mathrm{s} \rVert^2$.
Thus, we have
\begin{align}
    Z \sim \mathcal{CN}(\xi \mu_x,\xi^2 \sigma_x^2 + (1 - \xi^2) d_k^{-\tau} \lVert \mathbf{w}_\mathrm{s} \rVert^2).
    \label{z}
\end{align}
To compute the fourth moment of $Z$, we note that
if a complex random variable $S \sim \mathcal{CN}(\mu, \sigma^2)$, then we have (i) $S = \mu + W$, where $W \sim \mathcal{CN}(0, \sigma^2)$ (ii) $|S|^2 = |\mu + W|^2 = |\mu|^2 + |W|^2 + 2\, \mathrm{Re}(\mu^\H W)$. By using the distribution of $W$ along with some algebraic manipulations we get (iii)
$
\mathbb{E}[|S|^4]
  = |\mu|^4 + 4|\mu|^2 \sigma^2 + 2 \sigma^4$.
Finally, by utilizing these moments 
and \eqref{z}, we get \eqref{zk} and \eqref{zk_sq}.
\item \textit{Compute $\mathcal{X}_{3,k}$:}
We first compute the moments of $V_k$ as
  \begin{align}
  \label{vk1}
  \mathbb{E}[V_k]
     &= \mathbb{E}\Bigl[
         \sum_{n\in\mathcal{I}_a}
           |{h}_{\mathrm{r}k,n} |^{2}\,|\eta_n|^{2}
       \Bigr] \notag \\
     &= \sum_{n\in\mathcal{I}_a}
        \mathbb{E}\bigl[|{h}_{\mathrm{r}k,n} |^{2}\bigr]\,
        |\eta_n|^{2},  
        \end{align}
        \begin{align}
  \mathbb{E}[V_k^{2}] 
     &= \mathbb{E}\Bigl[
          \Bigl(
            \sum_{n\in\mathcal{I}_a}
              |{h}_{\mathrm{r}k,n} |^{2}\,|\eta_n|^{2}
          \Bigr)^{\!2}
        \Bigr] \nonumber \\
     &= \sum_{n\in\mathcal{I}_a}
          |\eta_n|^{4}\,
          \mathbb{E}\bigl[|{h}_{\mathrm{r}k,n} |^{4}\bigr] +2
        \sum_{\substack{n,m\in\mathcal{I}_a\\ n\neq m}}
          |\eta_n|^{2}\,|\eta_m|^{2}  \nonumber \\
     &\quad \times
          \mathbb{E}\bigl[|{h}_{\mathrm{r}k,n} |^{2}\bigr]\,
          \mathbb{E}\bigl[|{h}_{\mathrm{r}k,m} |^{2}\bigr]. 
    \label{vk2}\end{align}
Using the moments of $V_k$ from \eqref{vk1} and \eqref{vk2} and the scaled-Gamma property, we derive the statistics of \(\mathcal{X}_{3,k}\), stated in Lemma~\ref{lemma:RV}, which completes the proof.
\end{itemize}
\section{Proof of Theorem~\ref{thm:out_accurate}}
\renewcommand{\theequation}{B.\arabic{equation}}
\setcounter{equation}{0}
\label{app:outage}
{To obtain a tractable closed-form expression, $\mathcal{X}_{1,k}$, $\mathcal{X}_{2,k}$, and $\mathcal{X}_{3,k}$ are modeled as mutually independent under a product-form approximation. Since each effective channel entry is a sum of $N$ independently faded source--HRIS--user components, its marginal distribution can be approximated using the Gamma/moment-matching characterization in Lemma~\ref{lemma:RV}. Let $\mathbf{e}_\mathrm{s}=\mathbf{w}_\mathrm{s}/\|\mathbf{w}_\mathrm{s}\|$ and $q_k=\mathbf{e}_\mathrm{s}^{\H}\hat{\mathbf{h}}_k$. Then, $U_k=|q_k|^2+\|\hat{\mathbf{h}}_k-\mathbf{e}_\mathrm{s}q_k\|^2$, whereas $Z_k=\|\mathbf{w}_\mathrm{s}\|^2|q_k|^2$. Hence, although $U_k$ and $Z_k$ share $|q_k|^2$ and are not exactly independent, the remaining term in $U_k$ is obtained from the component orthogonal to the sensing direction. Under the isotropic Gaussian approximation, this component is independent of $q_k$~\cite[Appendix A.1.2]{tse2005fundamentals}. Moreover, since $V_k$ involves only the active subset $\mathcal{I}_\mathrm{a}$ with $N_\mathrm{a}\ll N$, $\mathcal{X}_{3,k}$ is also approximated as independent of the effective channel-power terms.}
From \eqref{out_3}, after some algebraic manipulations, we get
\begin{align}
\label{pout}
\mathcal{P}_{\mathrm{o}_k} 
  &= \mathbb{P}\left[
\frac{\mathcal{X}_{1,k}}{\mathcal{X}_{2,k} + \mathcal{X}_{3,k} + 1} \leq \gamma_k^{\mathrm{th}} 
     \right] \notag \\
  &= \mathbb{P}\left[
       {\mathcal{X}_{1,k}}-\gamma_k^{\mathrm{th}}({\mathcal{X}_{2,k} + \mathcal{X}_{3,k} }) \leq \gamma_k^{\mathrm{th}}
     \right].
\end{align}
Let $\mathcal{X}_{4,k}
	= \gamma_k^{\mathrm{th}}({\mathcal{X}_{2,k} + \mathcal{X}_{3,k} })
	\triangleq B$, which is the sum of two independent Gamma RVs.
Now, let $Y$ be an RV representing the sum of $L$ independent Gamma RVs, where each summand has shape and scale parameters $\alpha_\ell$ and $\beta_\ell$, respectively. Then, the probability density function (PDF) of $Y$ is given as \cite{ansari2012sum}
\begin{equation}
f_{Y}(y)
= 
{\prod_{\ell=1}^{L} (\beta_\ell)^{\alpha_\ell}}
\bar{\mathrm{H}}_{0, L}^{0, L}
  \left[
    e^y \;\middle|\;
    \begin{array}{c}
     \mathcal
    T_\ell^{(1)}  \\
    \mathcal
    T_\ell^{(2)} 
    \end{array}
  \right],
\label{pdf_foxh}
\end{equation}
where $y > 0$ and $    \mathcal
    T_\ell^{(1)} $ and $    \mathcal
    T_\ell^{(2)} $.
By using \eqref{pdf_foxh}, the PDF of $B$ can be written as
$ f_B(y)
    = \prod_{\ell=1}^2\beta_\ell^{\alpha_\ell}\,
      \bar{\mathrm{H}}_{0,2}^{2,0}\!\Bigl[
        e^{y}\;\Big|\;
        \bigl(\mathcal{T}_2^{(1)}\bigr),
        \bigl(\mathcal{T}_2^{(2)}\bigr)
      \Bigr]
\quad (y>0)$, where
$\alpha_1=\alpha_{Z,k}$ , $
\alpha_2=\alpha_{V,k}$, $
\beta_1=\frac{\beta_{Z,k}}{(\beta_k\alpha_\mathrm{s}\rho_\mathrm{t})}$, and $
\beta_2=\frac{\beta_{V,k}}{\tilde{\beta}_k}$.
Then, the PDF of $\mathcal{X}_{4,k}$ is given by the RV transformation as
\begin{equation}
  f_{\mathcal{X}_{4,k}}(y)
    = ({1}/{\gamma_k^{\text{th}}})
      f_{B}\left({y}/{\gamma_k^{\text{th}}}\right),
\quad (y>0). \label{PDF_X4}
\end{equation}
We next rewrite \eqref{pout} as follows
\begin{align}
\mathcal{P}_{o,k}
&=\mathbb{P}\bigl[\mathcal{X}_{1,k} - \mathcal{X}_{4,k} \le 0\bigr]
\nonumber\\ &=
\int_0^\infty
     F_{\mathcal{X}_{1,k}}(y+\gamma_k^{\text{th}})
     f_{\mathcal{X}_{4,k}}(y) dy.
     \label{pout_integral}
\end{align}
Substituting \eqref{PDF_X4}
and $F_{\mathcal{X}_{1,k}}$ from Lemma~\ref{lemma:RV} into \eqref{pout_integral} yields 
\begin{align}
\mathcal{P}_{o,k}
&= \int_{0}^{\infty}
 \frac{\gamma\left(
        \alpha_{\lVert\hat{\mathbf{h}}_k\rVert^{2}},
        \frac{\beta_{\lVert\hat{\mathbf{h}}_k\rVert^{2}}}{\rho_\mathrm{t} \beta_k}
        (y+\gamma_k^{\text{th}})
     \right)}
     {\Gamma\left(\alpha_{\lVert\hat{\mathbf{h}}_k\rVert^{2}}\right)}\prod_{\ell=1}^2{\,\beta_\ell^{\alpha_1}\beta_2^{\alpha_2}}
     \notag\\
&\quad\times \frac{1}{\gamma_k^{\text{th}}},
     \bar{\mathrm{H}}_{0,2}^{2,0}\!\Bigl[
       e^{y/\gamma_k^{\text{th}}}\;\Big|\;
       \mathcal{T}_2^{(1)},\mathcal{T}_2^{(2)}
     \Bigr] dy
     \label{integral_gcq}.
\end{align}
Since \eqref{integral_gcq} contains a Fox-${\mathrm{H}}$ function integral, direct evaluation is intractable. Hence, we apply the Gauss–Chebyshev quadrature (GCQ) rule \cite{singh2023performance} to obtain accurate OP expressions in Theorem~\ref{thm:out_accurate}. This completes the proof.

\section{Proof of Theorem~\ref{thm:rate_accurate}}
\renewcommand{\theequation}{C.\arabic{equation}}
\setcounter{equation}{0}
\label{app:rate}
The ER of the common message is defined as
\begin{align}
\mathcal{R}_{\mathrm{c},k}
   &\triangleq \mathbb{E}\bigl[\log_2(1+\gamma_{\mathrm{c},k})\bigr] \notag\\
   &= \frac{1}{\ln(2)}
      \int_{0}^{\infty}
        \frac{1-F_{\gamma_{\mathrm{c},k}}(x)}{1+x}{d}x,
\label{rate_define}
\end{align}
where $F_{\gamma_{\mathrm{c},k}}(x)$ is the CDF of the
$\gamma_{\mathrm{c},k}$.
Based on \eqref{sinrs} , we have \begin{align}F_{\gamma_{\mathrm{c},k}}(x)
  &= \mathbb{P}\bigl[\gamma_{\mathrm{c},k}\le x\bigr] \nonumber \\
  &= \mathbb{P}\left[
       \frac{\alpha_\mathrm{c}\rho_\mathrm{t}\beta_k U_k}
            {\alpha_k\rho_\mathrm{t}\beta_k U_k
             +\beta_k\alpha_\mathrm{s}\rho_\mathrm{t} Z_k + \tilde{\beta}_k V_k}
       \le x
     \right].
     \label{ratea1}
     \end{align}
With some algebraic manipulations in \eqref{ratea1}, we get
\begin{align}
 F_{\gamma_{\mathrm{c},k}}(x)
  = \mathbb{P}\left[
     \frac{ \rho_\mathrm{t}\beta_k U_k}{\bigl(\beta_k\alpha_\mathrm{s}\rho_\mathrm{t} Z_k + \tilde{\beta}_k V_k\bigr)}
      \le
      R_{1,k}
\right],   
\end{align}
where $R_{1,k} \triangleq\frac{x}{\alpha_\mathrm{c}-\alpha_k x}$.
Following the similar steps used in the outage analysis to obtain Theorem~\ref{thm:out_accurate}, we can obtain $ F_{\gamma_{\mathrm{c},k}}(x)$. We then apply the GCQ rule in \eqref{rate_define} to get 
\begin{equation}
\mathcal{R}_{\mathrm{c},k} \approx
\frac{2\pi}{T_2\ln(2)}
\sum_{i=1}^{T_2}
  \frac{\sqrt{1-y_i^{\,2}}}{1-y_i}\,
    \frac{1}{2}-
\left(\frac{F_{\gamma_{\mathrm{c},k}}\bigl(
      \tfrac{1+y_i}{1-y_i}
    \bigr)}{2}\right).
\label{rate_appendix}
\end{equation}
Finally, by substituting $ F_{\gamma_{\mathrm{c},k}}(x)$ into \eqref{rate_appendix}, we get $\mathcal{R}_{\mathrm{c},k}$. The ER for the private message can be computed by following the similar procedure as for $\mathcal{R}_{\mathrm{c},k}$. Thus, the unified expression for the ER of common and private messages is given in Theorem~\ref{thm:rate_accurate}. Here, we omit the detailed steps due to the space constraints. Finally, the ergodic sum rate can be obtained by using Theorem~\ref{thm:rate_accurate} in \eqref{rate_1}.

\section{
Proof of Theorem~\ref{thm:CRLB}}
\renewcommand{\theequation}{D.\arabic{equation}}
\setcounter{equation}{0}
\label{app:crlb}
From the FIM given in \eqref{fim}, we can write the CRLB as
\begin{align}
\mathrm{CRLB}(\theta)=  [\mathbf{F}]^{-1} &=\left[\mathbf{F}_{\theta \theta}-\mathbf{F}_{\theta \boldsymbol{\beta}_\mathrm{s}}\mathbf{F}^{-1}_{\boldsymbol{\beta}_\mathrm{s} \boldsymbol{\beta}_\mathrm{s}}\mathbf{F}^{\T}_{\theta \boldsymbol{\beta}_\mathrm{s}}\right]^{-1} \nonumber \\
&=\left[\mathbf{F}_{\theta \theta}- \frac{|\mathbf{F}_{\theta \boldsymbol{\beta}_\mathrm{s}}|^2}{\mathbf{F}_{\boldsymbol{\beta}_\mathrm{s} \boldsymbol{\beta}_\mathrm{s}}}\right]^{-1}.
\label{crlb1}
\end{align}
We then evaluate the terms required to compute the $\mathrm{CRLB}$ as
\begin{itemize}
    \item \textit{Compute $\mathbf{F}_{\theta \theta}$:} Based on \eqref{f_tt} we get
    \begin{align}
 &= \mathrm{tr} \left[ \left( \mathbf{b} \dot{\mathbf{a}}^\H + \dot{\mathbf{b}} \mathbf{a}^\H \right)
  \left( \dot{\mathbf{a}} \mathbf{b}^\H + \mathbf{a} \dot{\mathbf{b}}^\H \right) \mathbf{R}_\mathbf{x} \right] \nonumber \\
&= \mathrm{tr} \left[ \left( \mathbf{b} \dot{\mathbf{a}}^\H \dot{\mathbf{a}} \mathbf{b}^\H 
  + \mathbf{b} \dot{\mathbf{a}}^\H \mathbf{a} \dot{\mathbf{b}}^\H 
  + \dot{\mathbf{b}} \mathbf{a}^\H \dot{\mathbf{a}} \mathbf{b}^\H
  + \dot{\mathbf{b}} \mathbf{a}^\H \mathbf{a} \dot{\mathbf{b}}^\H\right) \mathbf{R}_\mathbf{x} \right] \nonumber \\
&\overset{(a)}{=}\mathrm{tr} \left[ \left( \|\dot{\mathbf{a}}\|^2 \mathbf{b} \mathbf{b}^\H + {N_\mathrm{t}}\, \dot{\mathbf{b}} \dot{\mathbf{b}}^\H \right) \mathbf{R}_\mathbf{x} \right] \nonumber\\
&= \|\dot{\mathbf{a}}\|^2 \,  \left( \mathbf{b}^\H \mathbf{R}_\mathbf{x} \mathbf{b} \right) 
  + {N_\mathrm{t}}\,  \left( \dot{\mathbf{b}}^\H \mathbf{R}_\mathbf{x} \dot{\mathbf{b}} \right)    ,
    \end{align}
where the equality $(a)$ follows because 
$\mathbf{a}^\H\dot{\mathbf{a}}=0$,
$\mathbf{b}^\H\dot{\mathbf{b}}=0$, and 
$\|\mathbf{a}\|^2={N_\mathrm{t}}$.
\item \textit{Compute $\mathbf{F}_{\theta \boldsymbol{\beta}_\mathrm{s}}$:} We note from \eqref{f_tb} that 
\begin{align}
\mathbf{F}_{\theta \boldsymbol{\beta}_\mathrm{s}}
&= \frac{2L}{\sigma_\mathrm{s}^2} \Re \left\{ \mathbf{a}^* \mathrm{tr} \left[ \mathbf{D}_{\mathrm{st}} \mathbf{R}_\mathrm{x} \dot{\mathbf{D}}_{\mathrm{st}}^\H \right] \right\} ,
\end{align}
where
\begin{align}
\mathrm{tr} \left( \mathbf{D}_{\mathrm{st}} \mathbf{R}_\mathrm{x} \dot{\mathbf{D}}_{\mathrm{st}}^\H \right) &= \mathrm{tr} \left( \dot{\mathbf{D}}_{\mathrm{st}}^\H \mathbf{D}_{\mathrm{st}} \mathbf{R}_\mathrm{x} \right)\nonumber  \\
&= \mathrm{tr} \left( (\dot{\mathbf{b}} \mathbf{a}^\H + \mathbf{b} \dot{\mathbf{a}}^\H)^\H (\mathbf{b} \mathbf{a}^\H) \mathbf{R}_\mathrm{x} \right) \nonumber \\
&= \mathrm{tr} \left[ (\mathbf{a} \dot{\mathbf{b}}^\H + \dot{\mathbf{a}} \mathbf{b}^\H)(\mathbf{b} \mathbf{a}^\H) \mathbf{R}_\mathrm{x} \right] \nonumber \\
&= \mathrm{tr} \left\{ \mathbf{a} \dot{\mathbf{b}}^\H \mathbf{b} \mathbf{a}^\H + \dot{\mathbf{a}} \mathbf{b}^\H \mathbf{b} \mathbf{a}^\H \right\} \mathbf{R}_\mathrm{x} \nonumber \\
&\overset{(a)}{=}
0 .
\label{crlb2}
\end{align}
Thus, by using \eqref{crlb2} into \eqref{crlb1}, we get the expression for $\mathrm{CRLB}(\theta)=  [\mathbf{F_{\theta \theta}}]^{-1} $ which is presented in Theorem~\ref{thm:CRLB}, which completes the proof.
\end{itemize}

\bibliographystyle{IEEEtran}
\bibliography{IEEEabrv,references}
\end{document}